\documentclass[traditabstract]{aa}
\usepackage{graphicx}
\usepackage{rotating,subfigure,amssymb,afterpage}
\usepackage{txfonts}
\usepackage{natbib}
\begin{document}
   \title{XMM-Newton study of six massive, X-ray luminous galaxy clusters
systems in the redshift range z = 0.25 to 0.5
}

   \author{H. B\"ohringer\inst{1,2}. G. Chon\inst{1},  R.S. Ellis\inst{3}, R. Barrena\inst{4,5},
           N. Laporte\inst{6}}


\offprints{hxb@mpe.mpg.de}

   \institute{$^1$ University Observatory, Ludwig-Maximilians-Universit\"at M\"unchen,
                  Scheinerstr. 1, 81679 M\"unchen, Germany.\\
              $^2$ Max-Planck-Institut f\"ur extraterrestrische Physik,
                   D-85748 Garching, Germany\\
              $^3$ University College, Gower St, London WC1E 6BT, United Kingdom\\
              $^4$Instituto de Astrofisica de Canarias, C/Via Lactea s/n, E-38205 
                      La Laguna, Tenerife,Spain\\
              $^5$ Universidad de La Laguna, Departamento di Astrofisica, E-38206 
                      La Laguna, Tenerife, Spain\\
              $^6$ Cavendish Laboratory, University of Cambridge, 19JJ Thomson Avenue, 
                      Cambridge CB3 0HA, United Kingdom\\      
 }

   \date{Submitted 25/2/22}

\abstract{Massive galaxy clusters are interesting astrophysical and cosmological study
objects, but are relatively rare. In the redshift range $z = 0.25$ to $0.5$ which is,
for example, a favourable region for gravitational lensing studies, about 100 such systems
are known. Most of them have been studied in X-rays. In this paper we study the six
remaining massive clusters in this redshift interval 
in the highly complete CLASSIX survey which have so far not been observed 
with sufficiently deep exposures in X-rays. With data from our new XMM-Newton observations
we characterise their structures, derive X-ray properties such as the X-ray luminosity and
intra-cluster medium temperature and estimate their gas and total masses. We find that one
cluster, RXCJ1230.7+3439, is dynamically young with three distinct substructures in 
the cluster outskirts and RXCJ1310.9+2157/RXCJ1310.4+2151
 is a double cluster system.
 Mass determination is difficult in the systems with 
substructure. We therefore discuss several methods of mass estimation including scaling
relations. In summary we find that five of the six study targets are indeed massive clusters 
as expected, while the last cluster RXCJ2116.2-0309 is a close projection of a distant and a
nearby cluster which has led to a previous overestimation of its mass. 
In the XMM-Newton observation fields we also find three low redshift clusters close 
to the targets which are also analysed and described here.
In the field of RXCJ2116.2-0309 we discover serendipitously a highly variable X-ray source which
has decreased its flux within a year by more than a factor of eight. This source is
most probably an AGN.  
}


\authorrunning{B\"ohringer et al.}
\titlerunning{Six massive galaxy clusters systems at z = 0.25 - 0.5}
   \maketitle
%

\section{Introduction}

Massive galaxy clusters are the largest well defined and
quasi-stable objects in our Universe and they constitute
interesting study objects and astrophysical laboratories.
They cause the largest deflections of light rays by general 
relativistic effects. This is successfully exploited in using
them as gravitational lens telescopes for the detailed study of
distant objects which would otherwise be too faint for such 
investigations, e.g. \citet{Eli2014,Bou2014,McL2016,Fuj2016}. 
They impose the strongest
modifications of the cosmic microwave background through the 
Sunyaev-Zeldovich effect, e.g. \citet{Sun1972}. 
Cluster mergers are the most energetic
events in the Universe after the big-bang, e.g. \citet{Sar2002},
and the shock waves in these mergers releasing an energy of
the order of $10^{64}$ erg are interesting large-scale sites
for cosmic ray acceleration which gives raise to the observed 
Mpc size radio halos and relics, 
e.g. \citet{Fer2012}. The most massive clusters also 
provide the most sensitive probes for the 
fluctuation amplitude of the cosmic matter density 
distribution and are thus important
test objects for cosmological models, 
e.g. \citet{Vik2009,Boe2014,Man2014}.

Therefore, there is a considerable interest in these clusters.
But these high mass systems are very rare.
Our {\sf CLASSIX} galaxy cluster survey of X-ray luminous
clusters in the nearby Universe out to redshifts of $\sim 0.5$
(\citet{Boe2016})
provide a good resource for looking for such objects, since
the survey is highly flux complete. It provides the advantage
that the X-ray luminosity is tightly related to the cluster 
mass, e.g. \citet{Vik2006,Pra2009}, and massive clusters
are therefore easily found among the most X-ray luminous systems.

Selecting the most X-ray luminous clusters 
with $L_X  \ge 6 \cdot 10^{44}$ erg s$^{-1}$  (0.1 to 2.4 keV),
we find more than a hundred clusters in the redshift range $z = 0.25 - 0.5$
in a survey volume of 15.7 Gpc$^3$, of which only five objects  
had no pointed X-ray observations so far
and one object has only a short XMM-Newton exposure. 
Due to the extremely large  interest in these systems, most of 
these clusters have been observed with XMM-Newton and/or Chandra. 
Here we report observations of the six remaining clusters with XMM-Newton,
which allow for a detailed study of their properties.

In the following, we describe the observations and the data reduction 
in section 2. Section 3 shows the observational results and outlines
the properties of the galaxy clusters. A discussion of the results and
implications on the structure and mass of the clusters is given in section 4,
followed by the conclusion in section 5.
In Appendix A we show the observed properties of the three low redshift
clusters which appear in the target fields and in Appendix B we 
describe a serendipitously detected highly variable X-ray source.   
For physical properties which depend on distance we use the following
cosmological parameters: $H_0 = 70$ km s$^{-1}$ Mpc$^{-1}$,
$\Omega_m = 0.3$ and a spatially flat metric.

\section{Observations and data reduction}

\begin{table}[h]
\caption{The six galaxy clusters systems studied in this paper, where
$z$ provides the cluster redshift.  
}
\begin{small}
\begin{center}
\begin{tabular}{cr}\hline
  name &   z \\ 
\hline
RXCJ1230.7+3439 & 0.3324 \\
RXCJ1310.9+2157 & 0.2781 \\
RXCJ1317.1-3821 & 0.2539 \\
RXCJ1414.6+2703 & 0.4770 \\
RXCJ1717.1+2931 & 0.2772 \\
RXCJ2116.2-0309 & 0.4390 \\
\hline
\end{tabular}
\end{center}
\end{small}
{{\bf Notes:} RXCJ1310.9+2157 is a double cluster system
including RXCJ1310.4+2151.}
\label{tab1}
\end{table}

Table~\ref{tab1} provides an overview of the observational targets.
The six clusters were observed with XMM-Newton in eight observations with the
observing IDs and exposure times given in Table~\ref{tab2}. The observations 
were conducted in our programme, except for the observation 
RXCJ1310.9+2157b, which was taken from the archive.
For two of the targets, RXCJ1310.9+2157 and RXCJ1717.1+2931 we also obtained
images with HST to use these cluster systems as gravitational lens telescopes.
The analysis of these observations is still ongoing.

\begin{table}[h]
\caption{XMM-Newton observations of the clusters. The observed and clean exposure
times are given in ks for the three detectors MOS1, MOS2 and PN, respectively.}
\begin{small}
\begin{center}
\begin{tabular}{lrrr}
\hline
  {\rm name} & {\rm XMM} &  {\rm obs.} & {\rm clean} \\
                   & {\rm ID}  &  {\rm time} & {\rm time}  \\
\hline
R..1230.7+3439  &0841900101& 27.7~27.8~22.2  & 27.6~27.7~20.7 \\
R..1310.9+2157a &0841900201& 11.4~11.4~~~8.4 & 11.2~11.2~~~8.1 \\
R..1310.9+2157b &0402250301& 15.8~15.8~10.2  & 15.6~15.5~10.1 \\
R..1317.1-3821  &0841900301& ~6.1~~~6.6~12.0 & ~0.3~~~0.6~~~3.2 \\
R..1414.6+2703  &0841900401& 22.3~22.3~17.8  & 21.5~21.4~17.1 \\
R..1717.1+2931  &0841900501& 27.9~28.2~21.3  & 19.3~16.7~~~0.7 \\
R..2116.2-0309a &0803410701& 21.8~31.9~26.4  & 16.4~27.9~~~4.0 \\
R..2116.2-0309b &0841900601& 31.6~31.6~25.5  & 21.8~21.9~14.4 \\
\hline
\end{tabular}
\end{center}
\end{small}
\label{tab2}
\end{table}

The data have been filtered in a two stage process by means of a 3$\sigma$-clipping 
technique, first in the hard band (10 - 12 keV for MOS and 10 - 14 keV for PN)
and then in a broad band (0.4 - 10 keV) to clean the data from times of enhanced
background due to solar flares. The original exposure times and the remaining clean
times are listed for each observation and detector in Table~\ref{tab2}.

For the further image and spectroscopic analyses, point sources were detected
by means of the XMM-Newton SAS source detection software with a combination of
eboxdetect, esplinemap, and emldetect. The detections were 
manually checked, spurious sources and substructure of the clusters 
were removed from the source list. Then these sources have 
been removed from the data for the spectroscopic and structural analysis.

For the image analysis, we corrected for the telescope vignetting by applying 
vignetting corrected exposure maps. To characterise the 
shape of the clusters and cluster components in an
approximate azimuthally symmetric way, we determined their surface brightness
profiles and fitted them with $\beta$-models \citep{Cav1976}. For multi-component
systems, the neighbouring components were cutout from the data in the analysis of 
each component.

For the spectroscopic analysis we follow in the first approach the method
of  \citet{Cho2017}. The spectral fitting was performed 
with the {\sf XSPEC} software \footnote{XPSEC is obtained from 
https://heasarc.gsfc.nasa.gov/xanadu/xspec/}. Two types of backgrounds 
were used in the analysis. For the particle background we used spectra
from filter wheel closed data adjusted to
the background level of the observation, by normalising with the ratio of the 
background rate in the un-illuminated corners of the detectors in
the target and filter wheel closed data.
The second background, the sky background, was obtained
by using three major sky background components (local bubble, hot halo 
and extragalactic X-ray background) fitted to the spectrum from a background region 
outside the cluster. For this we used a region outside a radius of at least 6 - 8
arcmin away from the cluster center, after
all point sources and non-relevant substructure were excised.
In addition to the three sky background components we also considered soft
protons, when necessary. The fit results
in the background were then used to fit the cluster and background 
model spectra to the spectra in the source region, with fixed parameters 
for the sky background. 

To get an approximate measure of the cluster masses, we assume spherical symmetry
of the clusters or cluster components and hydrostatic equilibrium. 
Simulations show, e.g. \citet{Bor2004,Nag2007,Ras2012},
that this may underestimate the cluster masses by about 10 - 30\%,  
which should be kept in mind. We do not correct for this 
in the values quoted for the hydrostatic mass determination.

We use the gas density profile of the intracluster medium as described by the beta
model obtained from the fit, to determine the cluster's gas mass and for the 
parameters to determine the cluster mass by means of 
the hydrostatic equation. The other information needed for the mass determination
is the plasma temperature profile. We do not have enough data to determine this profile very
accurately. Thus we use two bracketing solutions, which capture what is in general observed
for clusters in deep X-ray studies. As one extreme we use an isothermal profile and
as the other extreme we use a polytropic temperature profiles 
(where $T(r) \propto \rho^{\gamma-1}$) with a polytropic index up to $\gamma = 1.2$.
The observations show in general decreasing temperature profiles with a $\gamma$ less
than 1.2, e.g. \citet{Vik2006,Pra2007,Zha2008,DeG2004}. Instead of deriving a 
temperature profile in several radial rings, which is hardly possible with the given photon 
statistics, we determined the temperature when possible in several apertures and
often one outer ring, and check if the expected mean emission measure weighted 
temperature for a given polytropic index is consistent with the data.

\begin{figure}[h]
   \includegraphics[width=\columnwidth]{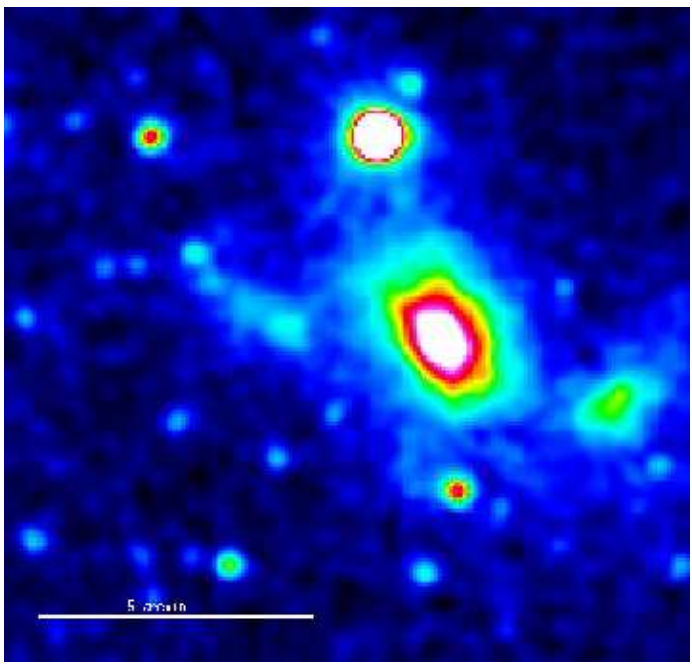}
\caption{XMM-Newton image of RXCJ1230.7+3439  in the 0.5 to 2 keV energy band
from all detectors combined smoothed with a Gaussian with $\sigma = 4$ arcsec.
A scale of 5 arcmin is indicated by the white bar. The image is color coded by 
the intensity of the surface brightness. 
}\label{fig1}
\end{figure}

\begin{figure}[h]
   \includegraphics[width=\columnwidth]{RXCJ1230+3439_ovl.ps}
   \includegraphics[width=\columnwidth]{R1230n_color.ps}
\caption{{\bf Top:}
Contours of the X-ray surface brightness (0.5 - 2 keV) overlayed 
on an optical PanSTARRS i-band image for the cluster RXCJ1230.7+3439. A black bar indicates
a scale of 2 arcmin.
Bottom: Color image of the cluster from the Legacy Imaging Surveys. 
The center (A), the south-west (B)
and eastern component (C) as well as the radio galaxy (D) are marked. 
}\label{fig2}
\end{figure}

The X-ray data are mostly reliable out to about $r_{500}$ 
\footnote{$r_{500}$ is the radius where the average
mass density inside reaches a value of 500 times the critical density
of the Universe at the epoch of observation.}.
Therefore we derive cluster mass estimates and intracluster gas masses for
this outer radius, quoted as $M_{500}$ and $M_{gas}$.

\section{Results}

\subsection{RXCJ1230.7+3439}

The cluster was found in the RASS by \citet{App1998} under the designation
ATZ98-D219 and by \citet{Boe2000} as part of the {\sf NORAS} survey, where its redshift 
was determined as $z = 0.3331$. The cluster was also detected via the  
Sunyaev-Zeldovich effect (SZE) in the Planck Survey
and is part of the second Planck cluster catalogue \citep{Pla2016}.
It was found in the Sloan Digital Sky Survey (SDSS) by \citet{Hao2010} 
and \citet{Wen2009}. It also appears in the Northern Sky Cluster
Survey \citep{Gal2003}. Table~\ref{tab3}
lists the position of the cluster and its substructures
determined from the X-ray maxima and several
cluster related redshifts.
 
Fig.~\ref{fig1} shows an XMM-Newton image of the cluster in the 0.5 to 2 keV 
energy band. We clearly see three diffuse X-ray components of the cluster:
The main cluster in the center, a subcomponent in the southwest and
diffuse emission in the east in the form of a spur. There is also a 
faint indication of a small extension of the cluster in the South.
In addition we see a few point sources in the field. 
The bright point source in the North of the 
cluster is associated to an NVSS radio galaxy, NVSS123050+344257, 
at the redshift of the cluster (z = 0.3335). A point source is also located
in the middle of the southern subcomponent, for which we have no identification.
 
We also performed an optical spectroscopic study of the system, which is 
described by \citet{Bar2021}. In this study 93 redshifts where determined in
the field of the cluster, of which 77 are cluster members. The mean 
redshift of the cluster members is $z=0.3324$ with a velocity dispersion of 
$\sigma_v = 1004 {+147 \atop122}$ km s$^{-1}$. A substructure analysis for the
galaxy distribution shows the same four substructure components mentioned
above. The mean redshifts of the three major substructures are given in  
Table~\ref{tab3}. The redshift of the southern component is $z= 0.3336$.
The structural analysis of the phase space distribution of the galaxies
indicates that the substructure is mainly oriented parallel to the
plane of the sky.
From the X-ray observations we can constrain the redshift of the intracluster 
plasma for the main component to be $z\sim 0.33$ and the eastern component
to be $z \sim 0.3$. Thus the different components form one dynamically young
system.

\begin{table}[h]
\caption{Properties of the components of the RXCJ1230.7+3439 cluster system.
}
\begin{small}
\begin{center}
\begin{tabular}{lllll}
\hline
  & {\rm Center} & {\rm Southwest} & {\rm East}\\
\hline
{\rm RA}        & 12~30~45    & 12~30~29.7    &12~30~59.5 \\
{\rm DEC}       &+34~39~07    &+34~37~50      &+34~39~21  \\
{\rm spec. z}   & 0.3324      &  0.3295       &  0.3306   \\
{\rm BCG z$^{a)}$}     & 0.3337      & 0.3269        & 0.3383  \\
{\rm X-ray~z}   & $0.33 \pm0.01$  & - &   $0.3 \pm0.05$ \\
$\sigma_v$      & 999 $\pm160$  & 792$\pm 230$ & $< 400$  \\ 
 F$_X$         &$7.2\pm 0.15$ &$2.4\pm 0.4$   &$1.2\pm 0.15$ \\
 L$_X$       & $2.4\pm 0.05 $ & $0.77\pm 0.14$ & $0.40\pm 0.05$ \\
 T$_X$       & $4.65  \pm 0.4$ & $4.4 \pm 0.6$  & $3.3 {+0.7\atop -0.6}$ \\
{\rm M$_{gas}$}   & 0.44       &  0.20         &  0.1      \\
 r$_c$          & 0.45        &  0.30         &  0.26    \\
 $\beta$        & 0.48        &  0.36         &  0.28    \\
\hline
\end{tabular}
\end{center}
\end{small}
{{\bf Notes:} The columns give the following parameters: RA and DEC are
for the epoch J2000,
spec. z is the mean of the spectroscopic redshifts of cluster galaxies
\citep{Bar2021}, BCG~z is the spectroscopic redshift of
the brightest cluster galaxy (BCG) at the center, X-ray~z is the redshift obtained from 
X-ray spectroscopy of the intracluster plasma, $\sigma_v$ is the velocity
dispersion from \citet{Bar2021}.  $F_X$ is the unabsorbed X-ray
flux in the 0.5 to 2 keV energy band in units of $10^{-13}$ erg s$^{-1}$ cm$^{-2}$,
$L_X$ is the k-corrected X-ray luminosity in the 0.5 to 2 keV energy band
in units of $10^{44}$ erg s$^{-1}$, $T_X$ is the mean temperature of the component
determined from X-ray spectroscopy in units of keV, $
M_{gas}$ is the gas mass inside $r_{500}$ in units of $10^{14}$ M$_{\odot}$,
$r_c$ is the core radius of the
intracluster medium in arcmin, and $\beta$ the parameter of the gas density profile slope.\\
${\bf ^{a)}}$ The first two BCG redshifts 
are from \citet{Bar2021} and they are
identical to those from the Sloan Digital Sky Survey (SDSS) Data release 13 (2016). 
The last BCG redshift is from SDSS.\\
}
\label{tab3}
\end{table}
\hspace{-8.0cm}

Fig.~\ref{fig2} shows an overlay of the contours of the X-ray surface brightness
on a PanSTARSS i-band image \footnote{PanSTARRS images can be retrieved from 
$https://ps1images.stsci.edu/cgi-bin/ps1cutouts$} and a colour composite 
image in the g, r, z band from the DESI Legacy Imaging Surveys \footnote{Legacy
Imaging Surveys data can be found at https://www.legacysurveys.org}. 
We clearly see BCG type galaxies
near the X-ray maxima in the three major substructure components. In the
south-western clump there are actually two bright galaxies near the X-ray
peak. The redshifts of these central galaxies are consistent with the mean
redshifts of the structures and the indicated X-ray redshifts of the 
intracluster medium (ICM).

Table~\ref{tab3} also provides information on the physical properties of the three main
substructures of RXCJ1230.7+3439 determined from the X-ray data.
The XMM-Newton observations provide good data for this cluster for all 
three detectors, with 27.6, 27.7 and 20.7 ks of clean exposure time for
MOS1, MOS2 and PN, respectively (see Table~\ref{tab2}). 
The X-ray image in the 0.5 to 2 keV energy band 
 shown in Fig.~\ref{fig1} was produced from the data of all three 
detectors with the appropriate scaling of the MOS exposure to that of PN.
The image is vignetting corrected and
background subtracted. In the imaging data we analysed the three major 
components of the cluster separately. We determined surface brightness 
profiles and $\beta$-model fits for each component. The results for the
fits are given in Table~\ref{tab3}. To assume an azimuthal 
symmetry for the eastern filament-like structure is a very crude approximation
which should only give a first impression. This should be kept in mind in 
interpreting the properties given for this cluster part. From the $\beta$-model
fits to the surface brightness profiles we can derive gas density profiles 
and total gas masses for the cluster components. Results for all 
three components are given in Table~\ref{tab3}.

From the X-ray data we determined average temperatures in six regions of the
cluster. For the central, main component we obtained temperatures in four regions,
$r \le 1$, $r \le 2$, $r \le 4$, $r = 2 - 4$ arcmin (corresponding
to $<286.5,~<573,~1146$ and $573 - 1146$ kpc), 
of, $4.6 \pm 0.3$, 
$4.8 \pm 0.3$, $4.47 \pm 0.3$ and $3.86 \pm 0.4$ keV, respectively. 
We explored which temperature profiles given by polytropic models
(of the form $T \propto \rho^{(\gamma-1)}$, where $\gamma$ is the polytropic 
index) are consistent with these measurements. For this aim we averaged
the model temperatures along the line of sight as a function
of projected radius applying the spectroscopic-like temperature method of \citet{Maz2004} 
and compared the results with the observations and their uncertainties.
We find that constant temperature and polytropic temperature profiles steeper
than $\gamma = 1.2$ can be ruled out with a significance of more than 1$\sigma$.
This justifies our approach of bracketing the temperature profile by these boundaries 
for the mass determination.
For the south-west component inside $r \le 1.6$ arcmin ($ \le 458$ kpc)
we find $T_x = 4.4 \pm 0.7$ keV  and for the eastern component, 
$T_X = 3.3 \pm 0.8$ keV at $r < 1$ arcmin ($< 286.5$ kpc). From the derived central cooling times
for the three substructures of, 14, 23, and 25 Gyr, respectively, we conclude
that none of the structures have a cooling core (where we assume a
cooling time of $< 5$ Gyr for a cooling core).

With the results for the intracluster plasma density and temperature distribution, 
we can determine first cluster mass estimates based on hydrostatic equilibrium,
which are given below in Table~\ref{tab9}. 
The error bars are the minimum and maximum of the mass values at $r_{500}$ 
derived from the different solutions for the temperature range given by the 
uncertainties and a constant temperature profile as well as polytropic model.
We discuss these results further in the next section.

\begin{figure}[h]
   \includegraphics[width=\columnwidth]{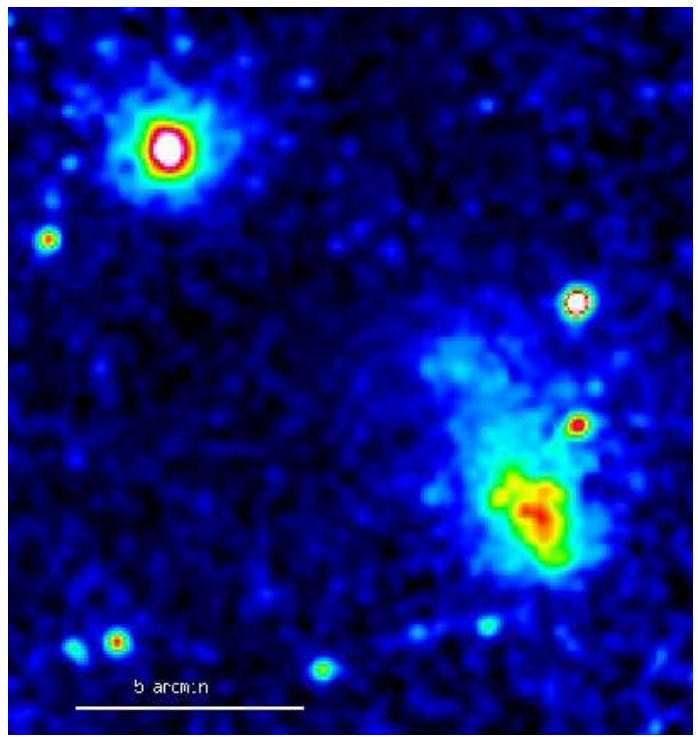}
\caption{XMM-Newton image of RXCJ1310.9+2157 (left) and RXCJ1310.4+2151 (right) 
in the 0.5 to 2 keV energy band. The white bar indicates a scale of 5 arcmin.
}\label{fig3}
\end{figure}

\begin{figure}[h]
   \includegraphics[width=\columnwidth]{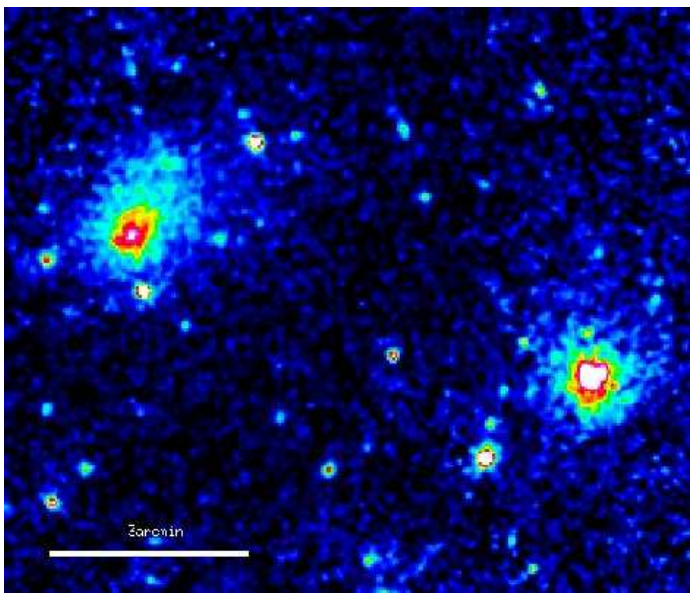}
\caption{XMM-Newton image of RXCJ1310.9+2157 (right) and 
RXCJ1311.7+2201 (left) in the 0.5 to 2 keV energy band.
The white bar indicates a scale of 3 arcmin.
}\label{fig4}
\end{figure}

\subsection{RXCJ1310.9+2157 and RXCJ1310.4+2151}

This system was found as double cluster in the {\sf NORAS II} survey \citep{Boe2017}. 
In one of the observations we see a third extended galaxy cluster X-ray source 
in the East of the two target clusters.
This cluster is located at lower redshift outside 
the redshift range for this cluster sample. We provide more
information of the properties of this cluster, RXCJ1311.7+2201, in the Appendix. 
RXCJ1310.9+2157 and RXCJ1310.4+2151 
have also been detected in the SDSS
\citep{Hao2010,Wen2012}, RXCJ1310.4+2151 also appears in \citet{Koe2007}. 
The sky position and redshifts of the two
system components are given in Table~\ref{tab4}. They have a similar 
distance and form a double cluster system. 

The components are covered by two XMM-Newton pointings which have  
the component RXCJ1310.9+2157 in common. X-ray images of the two 
pointings are shown in Figs~\ref{fig3} and \ref{fig4}. The cluster 
RXCJ1310.9+2157 in the East is
very round and compact, while RXCJ1310.4+2151 in the West
is slightly elongated and structured.

\begin{table}[h]
\caption{Cluster properties of RXCJ1310.9+2157 and RXCJ1310.4+2151}
\begin{small}
\begin{center}
\begin{tabular}{lcc}
\hline
{\rm name} & {\rm R..1310.9+2157} & {\rm R..1310.4+2151} \\
\hline
{\rm RA }       &  13~10~56.3     &    13~10~21.6  \\
{\rm DEC }      &  21~57~57.0     &    +21~50~04    \\
{\rm spec. z$^{a}$}   &  0.2781 {\rm (3)}  & 0.2734 {\rm (6)} \\
{\rm BCG z$^{b}$}   & 0.2822             & 0.2725    \\         
{\rm X-ray z} & $0.28\pm 0.03$     & $0.31\pm 0.04$  \\
\hline
 F$_X$       & $4.6 \pm 0.5$   & $7.1\pm 0.7$  \\
 L$_X$       & $1.1 \pm 0.1$  & $1.5 \pm 0.15$ \\
 T$_X$       & $4.75  \pm 0.5$ & $4.5 \pm 0.5$ \\
{\rm M$_{gas}$} & 3.0         &  4.3            \\
 r$_c$         &  0.19        & 1.15         \\
 $\beta$       &  0.48        & 0.75         \\ 
\hline
\end{tabular}
\end{center}
\end{small}
{{\bf Notes} The meaning of the parameters is the same as in Table~\ref{tab3}.\\
$^{a)}$ The integers in brackets give the number of galaxies available 
for the cluster redshift determination.\\ 
$^{b)}$ The redshift of the BCGs of RXCJ1310.9+2157  
is from the 2MASS galaxy survey \citep{Bil2014}. 
The one for RXCJ1310.4+2151 is from the SDSS Data release 13 (2016).
} 
\label{tab4}
\end{table}

\begin{figure}[h]
   \includegraphics[width=\columnwidth]{RXCJ1310.9+2157_ovl.ps}
   \includegraphics[width=\columnwidth]{R1310a_color.ps}
\caption{{\bf Top:} X-ray surface brightness contours overlayed on an optical 
PanSTARRS i-band image for RXCJ1310.9+2157.
{\bf Bottom:} Color image of the cluster from the Legacy Imaging Surveys 
}\label{fig5}
\end{figure}

\begin{figure}[h]
   \includegraphics[width=\columnwidth]{R1310+2151_DSS.ps}
   \includegraphics[width=\columnwidth]{R1310b_color.ps}
\caption{{\bf Top:} X-ray surface brightness contours overlayed on 
an optical DSS image for RXCJ1310.4+2151.
{\bf Bottom:} Color image of the cluster from the Legacy Imaging Surveys.
}\label{fig6}
\end{figure}

Figs.~\ref{fig5} and \ref{fig6} show overlays of contours of the X-ray surface 
brightness in the 0.5 to 2 keV band on optical images from PanSTARRS in 
the i-band for RXCJ1310.9+2157 and from DSS for RXCJ1310.4+2151. We also
show color composite images from the Legacy Imaging Surveys.
Both systems have clear BCG galaxies 
at their X-ray maxima. 
In RXCJ1310.4+2151 we note two further bright galaxies (B and C) which mark
the extension to the north.

From the analysis of the X-ray spectra of RXCJ1310.9+2157
in different circles and rings at radii,
$r < 0.5$, $< 1$, $< 2$ and $r = 1 - 2$ arcmin ($< 127, < 253, < 507, 253 - 507$ kpc),
, we find temperatures of
$T_X = 4.5 \pm 0.7$, $4.9 \pm 0.6$, $4.7 \pm 0.5$ and $4.5 \pm 0.8$ keV, respectively.  
For RXCJ1310.4+2151 we find for $r < 1$, $< 2.5$ and $r = 1 - 2.5$ arcmin
($ < 250, < 626, 250 - 626$ kpc) , temperatures of
$T_X = 4.0 \pm 0.6$, $4.5 \pm 0.5$ and $4.6 \pm 0.7$ keV, respectively. 
In both cases a temperature profile with a polytropic index of 1.2 or larger 
can be ruled out by the data. Cluster properties inferred from the 
observational data are given in Table~\ref{tab4}. The central cooling times
derived from the data of 6 and 28 Gyr for RXCJ1310.9+2157 and RXCJ1310.4+2151,
respectively, do not indicate a cooling core for the two systems.

\subsection{RXCJ1317.1-3821}

\begin{figure}[h]
   \includegraphics[width=\columnwidth]{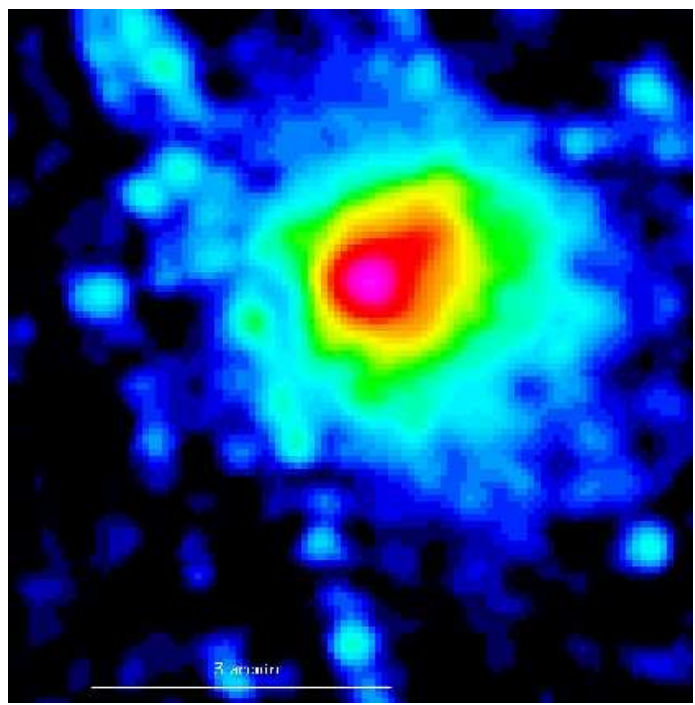}
\caption{XMM-Newton image of RXCJ1317.1-3821 in the 0.5 to 2 keV energy band.
A logarithmic scale was chosen for the color coding of the intensity. 
The white bar shows a scale of 3 arcmin.
}\label{fig8}
\end{figure}

RXCJ1317.1-3821 was identified in the course of the {\sf REFLEX} survey
\citep{Boe2004}
and the redshift was determined from two cluster members as $z=0.2772$.
The XMM-Newton observation is heavily affected by particles 
from a solar flare and only about 3 ks of usable pn data 
and hardly any MOS data could be 
recovered, not enough for a detailed analysis. Fig~\ref{fig8} shows an 
X-ray image of the center of the cluster from these data. The image 
is clearly affected by the gaps in the pn detector and the background noise.
An overlay of the X-ray contours on a DSS image \footnote{DSS (Digital Sky
Survey) images can be retrieved from: 
$http://www.sdss.org/dr13/data_access/bulk/$}
of the cluster are shown in
Fig.~\ref{fig9}. The cluster is very round and lobesided towards
the south-west. We clearly see the extended emission of the intracluster
medium of a galaxy cluster. A bright galaxy is found
at the X-ray maximum. The derivable properties of the cluster 
are given in Table~\ref{tab5}.

Due to the short exposure we do not see much of
the outer regions of the cluster. With the small number of photons
no spectroscopic analysis has been attempted. This study will be performed
with the new XMM-Newton and Chandra observations awarded to us. 

\begin{table}[h]
\caption{Cluster properties of RXCJ1317.1-3821
}
\label{t:log}
\begin{small}
\begin{center}
\begin{tabular}{lcc}
\hline
{\rm position}      & 13~17~13.0  & -38~21~57 \\
{\rm spec. redshift} & 0.2539     &  {\rm 2~gal.}  \\
{\rm BCG redshift$^{a}$}   & 0.2567     &              \\ 
\hline
 F$_X$       & $2.4 \pm 0.17$  \\
 L$_X$       & $4.2 \pm 0.3$   \\
\hline
\end{tabular}
\end{center}
\end{small}
{{\bf Notes} The meaning of the parameters is the same as in Table~\ref{tab3}
$^{a)}$ The redshifts of the BCG is from the {\sf REFLEX} survey \citep{Guz2009}.} 
\label{tab5}
\end{table}

\begin{figure}[h]
   \includegraphics[width=\columnwidth]{RXCJ1317_Opt_overlay.ps}
\caption{X-ray surface brightness contours overlayed on 
an optical DSS image for RXCJ1317.1-3821. 
}\label{fig9}
\end{figure}

\subsection{RXCJ1414.6+2703}

\begin{figure}[h]
   \includegraphics[width=\columnwidth]{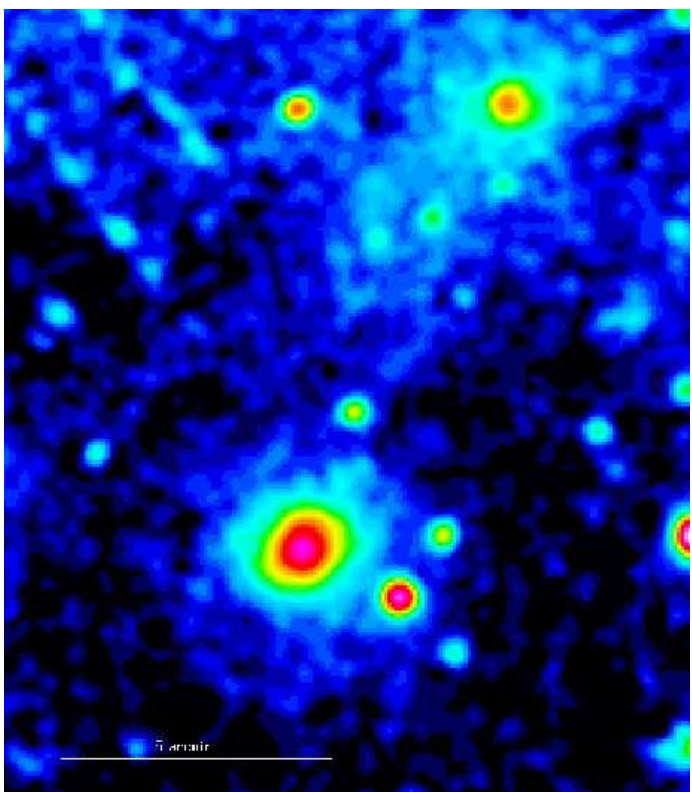}
\caption{XMM-Newton image of RXCJ1414.6+2703 in the 0.5 to 2 keV energy band.
The extended X-ray source in the North is the cluster RXCJ1414.3+2711 
(see Appendix A.2).
A scale of 5 arcmin is indicated by the white bar.
}\label{fig10}
\end{figure}

In the field of the cluster RXCJ1414.6+2703 we find two regions of 
extended X-ray emission. The X-ray source in the center of the 
field belongs to the cluster RXCJ1414.6+2703 (at $z = 0.4770$),
which was identified in the {\sf NORAS II} survey \citep{Boe2017}. 
The second source
in the north shows another clusters with the lable RXCJ1414.3+2711
at a different redshift of $z = 0.1619$. We describe this cluster in 
more detail in the Appendix.
RXCJ1414.6+2703 was also detected
in the SDSS \citep{Hao2010,Wen2010} and in the Planck Survey 
\citep{Pla2016}. 

From the analysis of the X-ray spectra of RXCJ1414.6+2703
in different circles and rings at radii,
$r < 0.4$, $< 1$, $< 2$ and $r = 1 - 2$ arcmin
($< 142, < 357, < 714, 357 - 714$ kpc),
we find temperatures of
$T_X = 7.0 \pm 0.7$, $7.1 \pm 0.5$, $6.8 \pm 0.45$ and $5.9 \pm 0.8$, respectively.  
The system shows a central cooling time
of 7 Gyr, which does not indicate a cooling core. 

\begin{table}[h]
\caption{Cluster properties of RXCJ1414.6+2703
}
\begin{small}
\begin{center}
\begin{tabular}{lc}\hline
                    & {\rm RXCJ1414.6+2703}  \\
\hline
{\rm RA}      & 14~14~39.0         \\
{\rm DEC}     &  +27~03~13          \\
{\rm spec. redshift} & 0.4770 {\rm (2)}   \\
{\rm BCG redshift$^{a)}$}   & 0.4772       \\
{\rm X-ray redshift} & 0.478 $\pm 0.02$ \\ 
\hline
 F$_X$         & $4.7 \pm 0.5$  \\
 L$_X$         & $3.4 \pm 0.4$  \\
 T$_X$         & $6.95 \pm 0.5$   \\
{\rm M$_{gas}$} & 6.8         \\
 r$_c$         &  0.25         \\
 $\beta$       &  0.54          \\ 
\hline
\end{tabular}
\end{center}
\end{small}
{{\bf Notes} The meaning of the parameters is the same as in Table~\ref{tab3}\\
$^{a)}$ The redshift of the BCG is from the SDSS data release 13 (2017). 
} 
\label{tab6}
\end{table}

\begin{figure}[h]
   \includegraphics[width=\columnwidth]{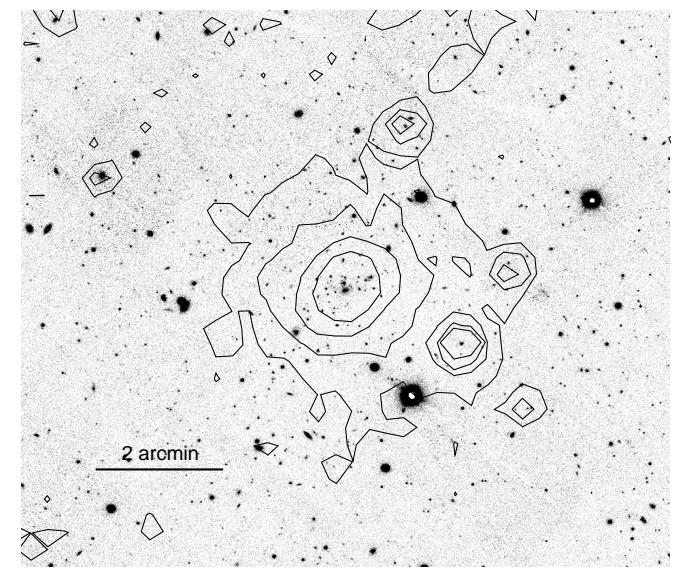}
   \includegraphics[width=\columnwidth]{R1414_color.ps}
\caption{{\bf Top:} X-ray surface brightness contours overlayed on 
an optical PanSTARRS i-band image for RXCJ1414.6+2703 
{\bf Bottom:} Color image of the cluster from the Legacy Imaging Surveys.
}\label{fig11}
\end{figure}

Fig.~\ref{fig11} presents an overlay of X-ray contours on an i-band image
obtained from PanSTARRS for RXCJ1414.6+2703 and a color image from the
Legacy Imaging Surveys. 
We note a round and compact 
cluster structure with a clear BCG at the X-ray peak. 

\subsection{RXCJ1717.1+2931}

\begin{figure}[h]
   \includegraphics[width=\columnwidth]{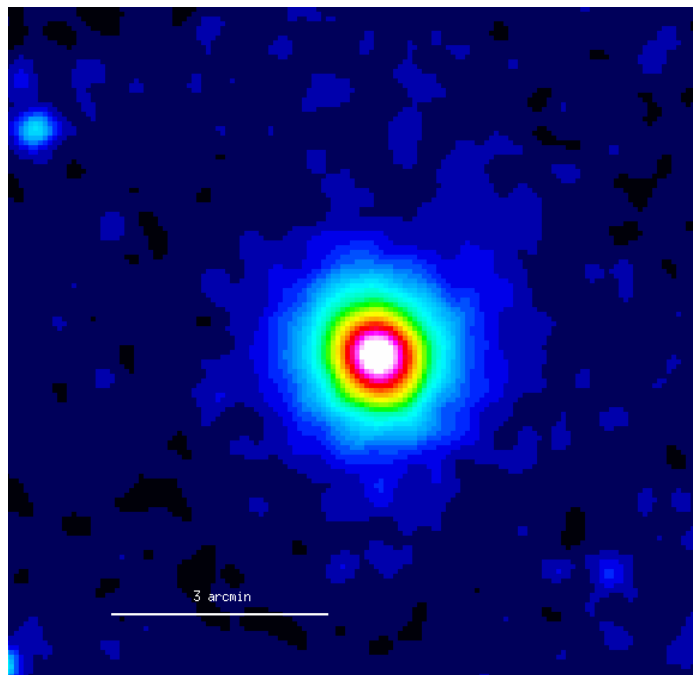}
\caption{XMM-Newton image of RXCJ1717.1+2931 in the 0.5 to 2 keV energy band.
A scale of 3 arcmin is shown by the white bar.
}\label{fig13}
\end{figure}

\begin{figure}[h]
   \includegraphics[width=6.5cm,angle=270]{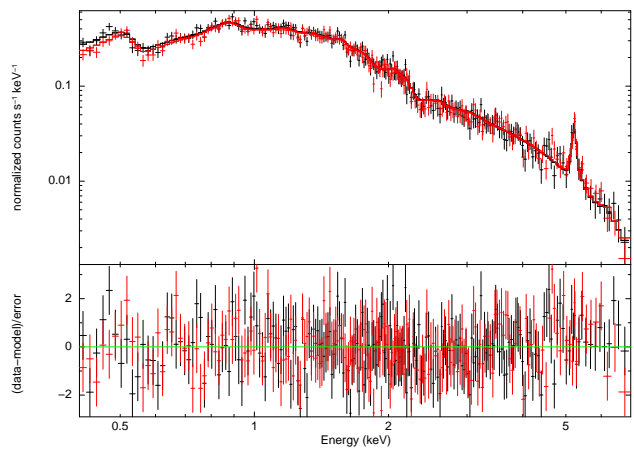}
\caption{XMM-Newton MOS spectrum of  RXCJ1717.1+2931 at $r \le 1$ arcmin.
The lines show fits of a model spectrum of hot thermal plasma at $\sim 4$ keV. 
}\label{fig13b}
\end{figure}

The cluster was discovered in the {\sf REFLEX II} survey \citep{Boe2013}.
\citet{Wen2015} found a cluster at this position in the SDSS
at $z=0.2768$. \citet{Fis1998} had previously identified the X-ray source as an AGN.
The XMM-Newton observation was affected by solar flare particles and the
pn data were effectively completely lost, with usable data of 19.3 ks and 16.7 ks 
exposure remaining for MOS 1 and 2, respectively. We therefore concentrate
here on a first X-ray analysis of the central region of the cluster.
A more detailed study will be performed with the new XMM-Newton and Chandra
observations awarded to us.

Fig.~\ref{fig13} shows an X-ray image in the 0.5 to 2 keV band produced from
the MOS data. RXCJ1717.1+2931 is a remarkably symmetric system 
and the X-ray emission is
highly peaked. The data remaining after flare cleaning are not sufficient
to unveil the temperature structure of the cluster spectroscopically.
From the MOS data we can obtain a spectrum of the central region, which is
shown in Fig.~\ref{fig13b}. The analysis of the spectrum yields a temperature
of $4 \pm 0.2$ keV and a redshift of $z = 0.275\pm 0.01$. The spectrum
can be well fit by a thermal spectrum at the cluster redshift. 
The clearly visible iron K-line
agrees perfectly with the line of highly ionised iron at the temperature 
and redshift of the cluster. It is not consistent with a fluorescent iron
line observed in some AGN. 

\begin{figure}[h]
   \includegraphics[width=\columnwidth]{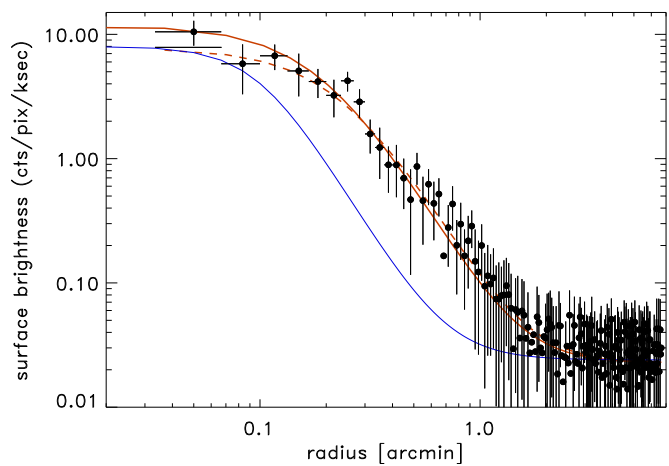}
\caption{X-ray surface brightness profile in the 0.5 to 2 keV band of
RXCJ1717.1+2931. The dashed red line shows the best fit of a $\beta$-model
convolved with the point spread function to the observed profile. The red
solid line shows the unconvloved profile and the blue solid line the profile
of a point source. 
}\label{fig14}
\end{figure}

A surface brightness profile of the X-ray emission of the cluster
is shown in Fig.~\ref{fig14}. The profile was fit by a $\beta$-model,
convolved with the MOS detector point spread function. We show in the 
Figure the convolved (dashed red line) and inferred unconvolved
profile (solid red line) and for comparison the profile 
of a point source (solid blue line).
The emission is clearly extended. Therefore, from the appearance of the
X-ray spectrum as well as from the source extent, we can safely conclude
that the X-ray emission of the cluster is not dominated by X-rays
from an AGN.

\begin{table}[h]
\caption{Cluster properties of RXCJ1717.1+2931
}
\label{t:log}
\begin{small}
\begin{center}
\begin{tabular}{lcc}\hline
\hline
{\rm position}       & 17~17~06.9  & +29~31~21 \\
{\rm spec. redshift} & 0.2772      &  {\rm 4~galaxies} \\
{\rm BCG redshift$^{a)}$}   & 0.2780      &               \\ 
{\rm X-ray redshift} & 0.275       & $\pm 0.01$  \\ 
\hline
 F$_X$         & $34 \pm 1$     &\\
 L$_X$         & $7.45 \pm 0.2$ &\\
 r$_c$         &  0.20          &\\
 $\beta$       &  0.68          &\\ 
\hline
\end{tabular}
\end{center}
\end{small}
{{\bf Notes} The meaning of the parameters is the same as in Table~\ref{tab3}.\\
$^{a)}$ The redshifts of the BCG is from the SDSS data release 2 (2004)
}
\label{tab7}
\end{table}

The cluster has a high X-ray luminosity and one would expect a higher
X-ray temperature from the scaling relation than what is measured.
We are most probably measure the lower central temperature of a cool core.
The analysis of the surface brightness profile indicates a cool core
with a central cooling time of about 1.2 Gyr and a cooling radius 
of about 100 kpc ($\sim 0.4$ arcmin). The X-ray luminosity in
the cooling region is about $0.5 \times 10^{44}$ erg s$^{-1}$.
A more detailed analysis will be possible with the future Chandra
and XMM-Newton data. 

Fig.~\ref{fig14b} shows X-ray surface brightness contours overlayed on an 
optical image from DSS. The X-ray peak coincides
with a bright BCG. A focus on the central region of the cluster is shown
in Fig.~\ref{fig14c}, where X-ray contours are superposed on an 
HST image taken with the F110 filter. The X-ray emission is precisely focused
on the central galaxy and we even note that the X-ray contours have a slight 
elliptical shape with a similar position angle as the BCG. Thus the X-ray trace
the gravitational potential of the galaxy, which is common for strong cool
core clusters.

\begin{figure}[h]
   \includegraphics[width=\columnwidth]{RXCJ1717_Opt_overlay.ps}
\caption{X-ray surface brightness contours overlayed on 
an optical DSS image for RXCJ1717.1+2931.
}\label{fig14b}
\end{figure}

\begin{figure}[h]
   \includegraphics[width=\columnwidth]{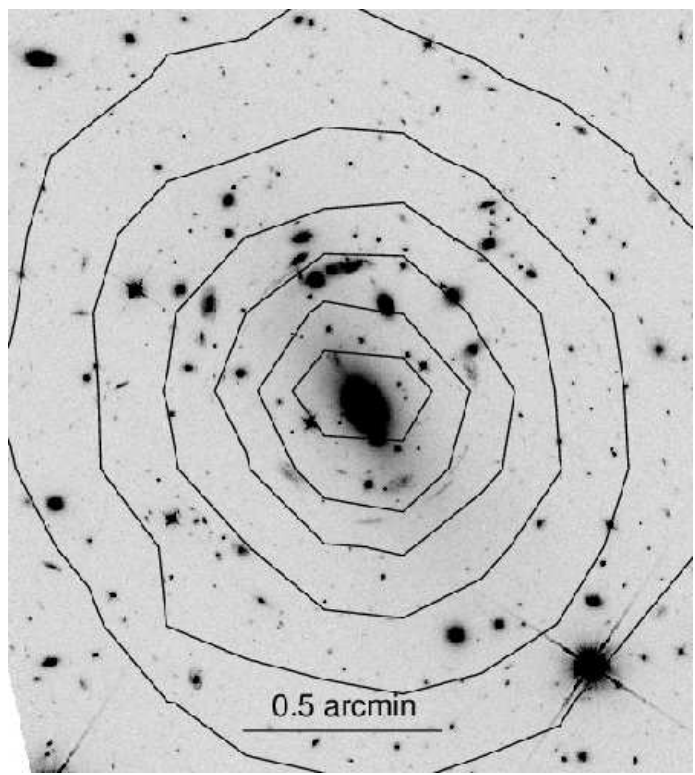}
\caption{Central region of the cluster RCJ1717.1+2931.
X-ray surface brightness contours are overlayed on an HST image.
}\label{fig14c}
\end{figure}

\subsection{RXCJ2116.2-0309}

The galaxy cluster RXCJ2116.2-0309
was identified in this field in the {\sf REFLEX II} survey
at a redshift of $z=0.4408$. In the ROSAT All-Sky Survey the cluster emission
is merged in a larger emission region with a foreground cluster,
RXCJ2116.1-0306, at redshift $z=0.2252$ and located to the north of the
distant target cluster. We describe the properties of the 
northern cluster in the Appendix. 

\begin{figure}[h]
   \includegraphics[width=\columnwidth]{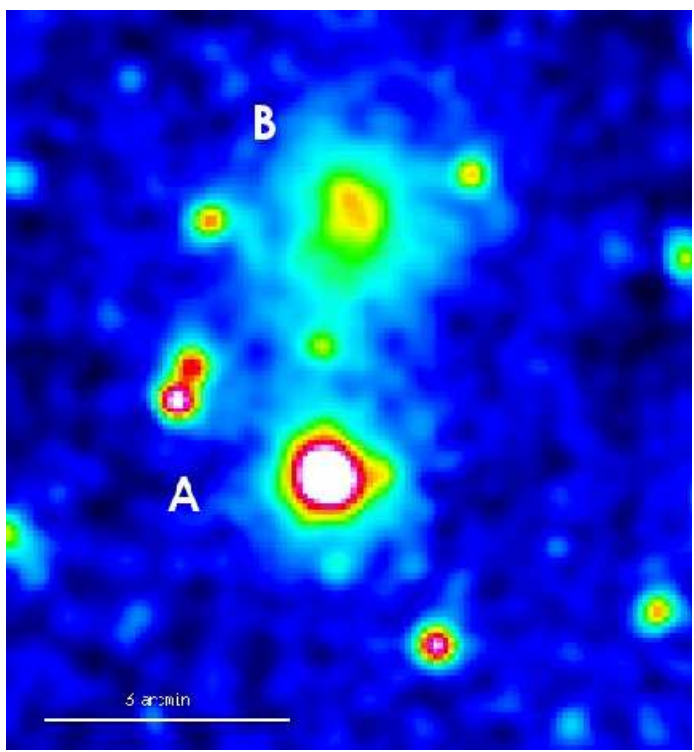}
\caption{XMM-Newton image of RXCJ2116.2-0309 in the 0.5 to 2 keV energy band.
This cluster is located in the middle of the image and labled A.
The extended emission in the north of the target cluster belongs to the 
cluster RXCJ2116.1-0306 labled B.
The white bar shows a scale of 3 arcmin.  
}\label{fig17}
\end{figure}

RXCJ2116.2-0309 was also found in the 
SDSS at similar redshifts by \citet{Wen2012} and \citet{Roz2015}.
Fig.~\ref{fig17} shows an XMM-Newton image of the two clusters.
RXCJ2116.2-0309 is very bright 
and compact and does not show much substructure. 
Table~~\ref{tab8} provides information on the redshift
and X-ray properties of the system.

\begin{figure}[h]
   \includegraphics[width=\columnwidth]{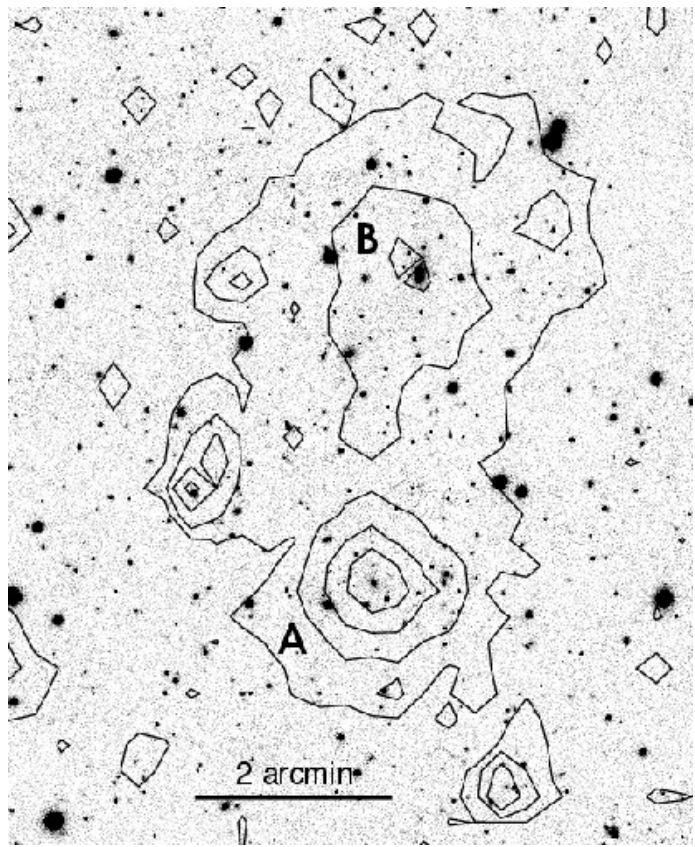}
\caption{X-ray surface brightness contours overlayed on 
an optical PanSTARRS i-band image for  RXCJ2116.2-0309 (A) 
and RXCJ2116.1-0306 (B).
}\label{fig18}
\end{figure}

\begin{figure}[h]
   \includegraphics[width=\columnwidth]{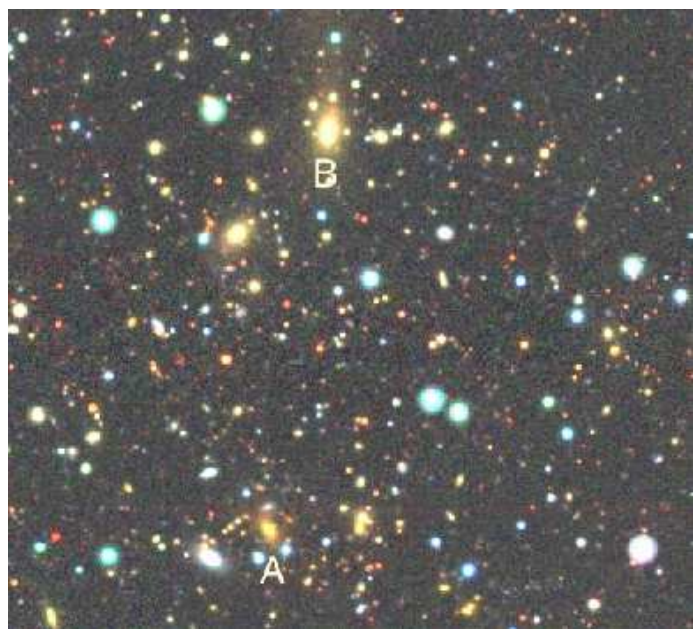}
\caption{ Color image of the clusters RXCJ2116.2-0309 (A) and RXCJ2116.1-0306 (B)
from the Legacy Imaging Surveys.
}\label{figX}
\end{figure}

Fig.~\ref{fig18} shows an image of the X-ray surface brightness 
contours overlayed on an optical i-band image from the
PanSTARRS survey and Fig.~\ref{figX}   
a colour image from the Legacy Imaging Surveys.
RXCJ2116.2-0309 features a clear BCG at the X-ray maximum.

\begin{table}[h]
\caption{Cluster properties of RXCJ2116.2-0309}
\label{t:log}
\begin{small}
\begin{center}
\begin{tabular}{lcccc}\hline
                    & {\rm RXCJ2116.2-0309} \\
\hline
{\rm RA}      & 21~16~12       \\
{\rm DEC}     & -03~09~26.5     \\
{\rm spec. redshift} & 0.4390 {\rm (1)} \\
{\rm BCG redshift$^{a)}$}   & 0.4390   \\
{\rm X-ray redshift} & 0.45 $\pm 0.015$  \\ 
\hline
 F$_X$         & $ 2.2 \pm 0.13$ \\
 L$_X$         & $ 1.4 \pm 0.4$ \\
 T$_X$         & $2.95 \pm 0.4$ \\
{\rm M$_{gas}$} & 1.8  \\
 r$_c$         & 0.12   \\
 $\beta$       & 0.51   \\ 
\hline
\end{tabular}
\end{center}
\end{small}
{{\bf Notes} The meaning of the parameters is the same as in Table~\ref{tab3}
$^{a)}$ The redshift of the BCG of RXCJ2116.2-0309 is from the SDSS data 
release 13 (2016).
}
\label{tab8}
\end{table}

\begin{table*}[h]
\caption{Results of different mass estimates for all the observed galaxy clusters described in the main text.}
\begin{small}
\begin{center}
\begin{tabular}{lrrrrrrr}
\hline
  {\rm mass } & {\rm 1230C} &  {\rm 1230SW} & {\rm 1230E} & {\rm 1310C}& {\rm 1310S} 
   & {\rm 1414S} & {\rm 2116S}  \\
\hline
$M_{500}(\beta)$    & 1.9$\pm0.4$ & 1.14$\pm0.2$ &  - & 2.0$\pm0.4$  & 3.3$\pm0.7$  & 3.9$\pm1.0$  & 1.02$\pm0.3$ \\
$M_{500}(\beta_{fix})$ & 3.1$\pm0.6$  & 2.9$\pm 0.5$ & - & 3.1$\pm0.6$ & 2.7$\pm0.6$ & 5.2$\pm1.5$ & 1.5$\pm0.4$  \\
$M_{500}(L_X)$     & 4.7$\pm1.9$ & 2.3$\pm0.9$ & 1.5$\pm0.6$ & 2.8$\pm1.1$ & 3.6$\pm1.4$ & 5.4$\pm2.1$ & 3.2$\pm1.3$ \\
$M_{500}(T_X)$     & 3.1$\pm0.4$ & 2.9$\pm0.16$  &  -    & 3.3$\pm0.5$ & 3.05$\pm0.5$  & 5.2$\pm0.6$ & 1.48 $\pm0.3$ \\
$M_{500}(M_{gas})$ & 4.2$\pm0.6$ & 2.3$\pm0.35$ & 1.2$\pm0.2$ & 3.05$\pm0.4$ & 4.1$\pm0.6$ & 6.0$\pm0.9$ & 2.0$\pm0.3$ \\
$M_{500}(Y_X)$    &  3.7$\pm0.4$  & 2.3$\pm0.25$ &  -    & 3.05$\pm0.3$ & 3.6$\pm0.4$  & 5.7$\pm0.6$ & 1.7$\pm0.2$  \\
$M_{dyn}$         & 5.6$\pm1.0$  &  2.7$\pm2.0$ & $\sim 0.7$ &  &     &      &        \\
\hline
$M_{500}$        & 3.7$\pm0.4$  & 2.5$\pm0.25$ & 1.35$\pm0.3$ & 3.1$\pm0.3$ & 3.4$\pm0.4$ & 5.5$\pm0.6$  & 1.9$\pm0.2$  \\
$r_{500}$        & 3.45 & 2.98 & 2.39  & 3.69 & 3.94 & 2.97   & 2.27 \\
\end{tabular}
\end{center}
\end{small}
{{\bf Notes:} All masses are given in units of $10^{14}$ M$_{\odot}$. $M_{500}(\beta)$ and $M_{500}(\beta_{fix})$ are hydrostatic mass 
estimates based on $\beta$-models for the surface brightness distribution of the intracluster medium, where the first value is based on
the fitted value for $\beta$ and the second for a model with $\beta = 2/3$. $M_{500}(L_X)$ and $M_{500}(T_X)$ are based on the X-ray 
luminosity - mass relation and temperature - mass relation, respectively. $M_{500}(M_{gas})$ is determined from the total gas mass 
in the cluster and the relation of the baryon fraction with cluster mass, $M_{500}(Y_X)$ is derived from the scaling relation 
of the product of gas mass and temperature with cluster mass, and $M_{500}$ is derived from the mean of the scaling relation results.}
\label{tab9}
\end{table*}

Intracluster plasma temperatures were determined from an X-ray spectroscopic
analysis. For RXCJ2116.2-0309, we determined temperatures in four regions,
with $r < 0.4$, $< 0.83$, $< 1.5$, and $r = 0.5 - 1$ arcmin 
($ < 136, < 283, < 511, 170 - 341$ kpc), yielding the
following results: $2.9\pm 0.5$, $3.0\pm 0.4$, $2.95\pm0.4$, 
and $2.60\pm 0.7$ keV, respectively. 
The system shows a central cooling time of the ICM of about 3 Gyr, which can be 
taken as a sign of a moderate cooling core.

In the field of the cluster we also observe an X-ray point source
which decreased its flux dramatically between the first and second pointed
observation. It is located at RA = 21 16 06.6, DEC = -03 11 32 and we thus
lable it as XMMUJ2116.2-0311. The source can be seen in 
Figs.~\ref{fig17} and \ref{fig18} about 2.5 arcmin  
south-west of RXCJ2116.2-0309. The observation shown is 0841900601
where the source is less bright. We have not found a previous identification of 
this source. It is most probably an AGN.
We describe further details of the X-ray properties of this source
in the Appendix.

\section{Discussion}

Some of the clusters in this sample for which we have
good X-ray data show a dynamically unrelaxed configuration. For such clusters
the mass determination is more difficult and uncertain than for relaxed
clusters. Therefore we went through a critical inspection of various
ways of mass estimation to determine the best overall consistent picture.
This is described in this section.

In addition to the hydrostatic mass determination
we used several well calibrated scaling relations to obtain mass estimates
from different direct and indirect observables. Based on the studies 
by \citet{Pra2009} and \citet{Vik2006} we adopted the following 
X-ray luminosity - mass relation:

\begin{equation} 
M_{500} = 3.25~  L_X^{1/3}~  (h_0/0.7)^{2/3} ~~~,
\end{equation}

where $M_{500}$ is the mass inside $r_{500}$ in units of $10^{14}$ M$_{\odot}$,
$L_X$ is the X-ray luminosity in the 0.5 - 2 keV band inside $r_{500}$,
and $h_0$ is the Hubble constant in units of 100 km s$^{-1}$ Mpc$^{-1}$.
For the uncertainty we adopted the scatter in the $L_X - M$ relation
of about 40\%.
For the mass temperature relation we use \citep{Arn2005}: 

\begin{equation}
M_{500} = 4.1~ {T_X \over 5{\rm keV}}^{1.49}~ E(z)^{-1}  
\end{equation}

where $T_x$ is the spectroscopic temperature inside $r_{500}$ and 
$E(z) = H(z)/H_0$. 
The main contribution to the uncertainty is in our case 
the error in the temperature measurement.

Further good proxies for the total cluster mass are the total
gas mass inside $r_{500}$, $M_{gas}$, and 
the parameter $Y_X$, which is the product of the
gas mass and the temperature. The latter is thought to be proportional
to the observable decrement in the microwave background through
the Sunyaev-Zeldovich effect \citep{Kra2006,Nag2006}. 
To scale the gas mass fraction, $f_{gas}$,  
with cluster mass we apply the observationally determined relation 
from \citet{Pra2009}.

\begin{equation}
f_{gas} = 0.09 \times \left({M_{500} \over 2 \times 10^{14} M_{\odot}}\right)^{0.21}~~.
\end{equation}

The uncertainty of $M_{500}(M_{gas}) = M_{gas}/f_{gas}$ is determined by adding the uncertainty
in the $\beta$-model fit taken from the difference between different detectors
and a 13\% uncertainty from the scatter of
the gas mass fraction - mass relation \citep{Pra2009}.
For the scaling relation of $Y_X$ with cluster mass we use \citep{Pra2009}:

\begin{equation}
M_{500} = 3.90~ {Y_X \over 2}^{0.561}~ E(z)^{-0.4}  
\end{equation}

where $Y_X$ is in units of $10^{14}$ M$_{\odot}\times$ keV. To determine
the relative uncertainty of this parameter we apply a Gaussian addition of
the error contributions from $\Delta T_X$ and $\Delta M_{gas}$. This may
be an overestimate since  $T_X$ and $M_{gas}$ are thought to be anti-correlated 
(e.g. \citet{Kra2006}). The saller influence of $\Delta T_X$ is, except for two 
cases in the Appendix, always less then 20\% and therefore the possible
overestimate is also small.
The results of the mass estimates are given in Table~\ref{tab9}.

\subsection{RXCJ1230.7+3439}

We note in Table~\ref{tab9} that the mass estimates obtained for 
the main component of RXCJ1230.7+3439 from
the different observables span a range from $M_{500} = 3.1 $ to $ 4.7 \times 10^{14}$ M$_{\odot}$.
The mass determined from the plasma density distribution and
the isothermal or polytropic temperature model is with a value of 
$M_{500}(\beta) = 1.9 \times 10^{14}$ M$_{\odot}$ significantly smaller than these values.

An explanation for this discrepancy is certainly the disturbed state of the clusters. A major reason
for the low resulting cluster mass using the $\beta$-model for the plasma density distribution
is the low value of $\beta$ derived from the observed surface brightness profile. 
Massive relaxed clusters show typical $\beta$-values of $2/3$ or even higher (e.g. \citet{Cro2008}).
Thus a low value of $\beta$ is most probably another manifestation that the hot plasma in
the cluster has not settled. Two factors can play a role here, on one side substructure 
in the line of sight outside $r_{500}$ projected onto the cluster, and inhomogeneities and
deviations from an azimuthally symmetric plasma distribution which boost the emission measure. 
To get a feeling for the effect of low $\beta$-values, we also calculated 
the cluster mass by fixing the slope parameter $\beta$ to $2/3$, while keeping all
the other model parameters the same. The result is shown in Table~\ref{tab9} under the
lable $M_{500}(\beta_{fix})$. With a mass value of $3.1 \times 10^{14}$ M$_{\odot}$, we now
get a value in the range of the results obtained from scaling relations, which seems
to support our reasoning. Since the $\beta$-values for most of the other unrelaxed clusters
studied here in detail are also low, we performed the same exercise with a fixed
value of $\beta$ also for the other systems.

Among the mass estimates $M_{500}(T_X)$ is on the low side for the
main component of RXCJ1230.7+3439. This is not 
unusual for unrelaxed clusters. For example, the study by \citet{Cho2017}
showed that unrelaxed clusters have temperatures below
the general scaling relation.

As a final value for the mass, we adopted an average result of the four scaling 
relations with $L_X$, $T_X$, $M_{gas}$, $Y_X$ and of $M_{500}(\beta_{fix})$. Among 
the mass proxies from scaling relations, the high value of $M_{500}(L_X)$ is
probably the most uncertain. This is due to the large scatter in the $L_X - M$
relation and in this unrelaxed cluster we may expect some projected substructure
outside $r_{500}$ projected onto the main cluster component. On the other hand
the parameters $T_X$ and $M_{gas}$ are quite reliable proxies. We account 
for this by using different weights for the parameters in the averaging process.   
For $M_{500}(L_X)$ the weight is 0.75, while for all other parameters the weight 
used is 1. Thus the weight for $M_{500}(T_X)$ and  $M_{500}(M_{gas})$ is effectively
1.5 as $M_{500}(Y_X)$ depends on both $T_X$ and $M_{gas}$.
For the uncertainty of the combined
mass result we use the relative error of $M_{500}(Y_X)$. The same procedure is
also used for the other clusters.

For the second, south-west component of RXCJ1230.7+3439 we note in Table~\ref{tab9}
quite consistent values for the mass from the observables and the model with
the fixed $\beta$-value in the range of 2.3 - 2.9 keV. Here one can question if
this component has been affected by extra heat from the interaction with the
main component, which would result in a temperature that is higher than what
corresponds to the mass of the object.  But since the mass obtained from the 
temperature as observable  is not much larger than the values from the other
proxies and since this component is separated from the main body, this effect
cannot be very large. 

For the eastern component the effect of interaction heating is probably more 
serious. This would certainly effect the temperature of 3.3 keV as a mass 
proxy. Using the observed X-ray luminosity and the $L_X - T_X$ relation,
we would predict a temperature of only 2.45 keV, supporting the idea
that the cluster component is heated by the interaction.
Since the X-ray surface brightness distribution of this component 
is quite irregular and cannot easily be modeled, a hydrodynamical mass 
model seems not appropriate. Therefore
we only use the X-ray luminosity and the gas mass as a tentative proxy
for the mass of this cluster part.

Finally we can compare the 
X-ray determined mass of this system to the dynamical mass determined
from the galaxy velocities by \citet{Bar2021}. For the main component the 
dynamical mass with a value of $M_{dyn~500} = 5.6 (\pm 1.0) \times 10^{14}$ M$_{\odot}$ is higher by
about 40\% than the values obtained from the X-ray observation. This can again
be attributed to the unrelaxed state of the cluster. In the study by \citet{Sar2013}
based on numerical cosmological simulations they find a scatter in the correlation
of $M_{dyn}$ and $M_{true}$ of about 30-40\%, where $M_{dyn}$ has been determined
in a way mimicking real observations. Since we expect to find very unrelaxed clusters 
in the high part of the scatter, this discrepancy is not too surprising. 
A more important reason for the different results may, however, come from the fact, that
the galaxies with redshifts cover a larger region on the sky than the aperture
with a radius of $r_{500}$ used in the X-ray study. We therefore determined
the velocity dispersion also for the 47 galaxies with redshifts inside $r_{500}$
and find a lower velocity dispersion of $\sigma_v = 913 \pm 133$ km s$^{-1}$,
which points to a mass of about  $M_{dyn~500} = 4.3 \times 10^{14}$ M$_{\odot}$,
in better agreement with the mass estimate from the X-ray data.

For
the south-western component $M_{dyn}$ and the mass from X-ray observations are
similar. Also for the eastern component the results for the X-ray and 
dynamical mass are close. 

In summary, we confirm that RXCJ1230.7+3439 is a massive, dynamically 
young cluster system, with a core region with a mass around  
$M_{500} = 4 \times 10^{14}$ M$_{\odot}$. The sum of the components
leads to a total mass of the system of 
$M_{500} = 7.7 (\pm 0.7) \times 10^{14}$ M$_{\odot}$ from the X-ray 
data and $M_{500} = 9.0 (\pm 2.3) \times 10^{14}$ M$_{\odot}$
from the optical data.

\subsection{RXCJ1310.9+2157 and RXCJ1310.4+2151}

For the cluster RXCJ1310.9+2157 we find masses of $M_{500} = 2.8 - 3.3 \times 10^{14}$  
M$_{\odot}$ for the different mass estimates shown in Table~\ref{tab9},
excluding the result for the low $\beta$ value. Thus we have results which
are all consistent with the finally adopted value of the mass of
$M_{500} = 3.1 (\pm 0.4) \times 10^{14}$  M$_{\odot}$. The structure of
the cluster is less complex as in the case of RXCJ1230.7+3439
and therefore the result is more clear.

RXCJ1310.4+2151 is elongated and has more substructure. Here we 
find a larger range of resulting masses, 
$M_{500} = 2.7 - 4.1 \times 10^{14}$ M$_{\odot}$. The mass estimates
involving the intracluster medium temperature are lower than those
derived from X-ray luminosity and gas mass. This indicates that 
the virialised part of the cluster is smaller than the system
contributing to the total X-ray luminosity and gas mass. In other
words the cluster is not fully virialsed, as can be expected
from the asymmetric X-ray appearance. The finally adopted total
mass of $M_{500} = 3.4 (\pm 0.5) \times 10^{14}$  M$_{\odot}$ may rather be
slightly underestimated.

\subsection{RXCJ1414.6+2703}

RXCJ1414.6+2703 appears very compact and symmetric. Again we observe that
the different mass estimates deliver consistent results. We note
a mass range of the estimates of $M_{500} = 5.2 - 6.0 \times 10^{14}$  
M$_{\odot}$, which amounts to only $\pm 7 \%$. The values are 
well consistent with the finally adopted result of
$M_{500} = 5.5 (\pm 0.7)\times 10^{14}$  M$_{\odot}$.

\subsection{RXCJ2116.2-0309}

In this case we have a partial overlap of two clusters in the line-of-sight.
This complicates the analysis. For the southern, distant target cluster we 
obtain mass estimates 
in the range of $M_{500} = 1.48 - 3.2 \times 10^{14}$  M$_{\odot}$. The highest
mass estimate comes from the X-ray luminosity as proxy. This may be affected by
projection effects as well as the second largest value from the gas mass proxy.
The temperature on the other hand yields a low value. As a finally
adopted value we get $M_{500} =  1.9 (\pm 0.4)\times 10^{14}$  M$_{\odot}$,
where we have increased the error from a value we obtained with our 
general scheme of $\pm 0.3$ to $\pm 0.4$ M$_{\odot}$, accounting
for the projection effects. The final value is thus consistent with all
estimates except for the X-ray luminosity proxy.

\section{Summary and Conclusions}

A major goal of the paper is to provide reliable mass estimates of the target clusters
of the XMM-Newton observations, as a reference for further studies, in particular
for using them as gravitational lense telescopes. They were selected
with the expectation to have masses larger than $M_{500} = 5 \times 10^{14}$ M$_{\odot}$.
Sufficient data for a structural study and mass determination were only available for
four of the targets, while for two targets we have to wait for the new XMM-Newton
and Chandra observations awarded to us. 

Some of the four well observed cluster systems show
either pronounced substructure or are double cluster systems. Because of this 
complication of dynamically young systems, we performed a critical inspection
of several different methods of mass determination to obtain a consistent
and reliable picture of the mass estimate. In particular we identify the
low value of the slope of the X-ray surface brightness profile as one 
of the signs of substructure in the outskirts of the cluster, which has to
be taken with care, if the data are not sufficient to perform a detailed
substructure analysis to remove all substructural features. 
Therefore we also modeled the clusters with $\beta$-parameters
fixed to a standard value of 2/3. With this approach we obtained consistent
results together with mass estimates from several well tested X-ray observables
which can be used as mass proxies.   

From these results we conclude that all, but one of the 
four systems satisfy the condition that the overall mass is larger
than  $M_{500} = 5 \times 10^{14}$ M$_{\odot}$. For RXCJ1230.7+3439 this is true
if all substructure components are added. This is also confirmed by the determination
of the dynamical mass of the cluster from the velocity dispersion of the galaxies.
For RXCJ1310.9+2157 and RXCJ1310.4+2151 the mass is larger than the limit considered 
when both components of the double system are added. For RXCJ1414.6+2703
we get a mass estimate of $M_{500} = 5.5 (\pm 0.7) \times 10^{14}$ M$_{\odot}$ 
for the compact and round cluster system. 
For RXCJ1317.1-3821 and RXCJ1717.1+2931the high observed X-ray luminosity
indicates massive clusters. But a firm conclusion has to await the analysis
of the upcoming observations.
Only for RXCJ2116.2+0309 the cluster system is not as massive as expected. In
the ROSAT All Sky Survey the X-ray emission of both cluster components is
merged into one emission region and the cluster mass was overestimated by
attributing the total luminosity to the distant cluster. 

The two clusters, for which we still await better X-ray data are both
more regular. RXCJ1717.1+2931 is round and compact with a pronounced
central surface brightness peak. It definitely has a strong cooling core.
The upcoming Chandra observation will be especially useful to unveil
the cool core structure. RXCJ1317.1-3821 is not perfectly symmetric and has
some interesting structural features in the centre which also will be
uncovered in more detail with the future Chandra observation.

\begin{acknowledgements}  
We thank the anonymous referee for helpful comments.
G.C. acknowledges support by the Deutsches Luft- und Raumfahrtzentrum
under grant no. 50 OR 1905. H.B. acknowledges 
support from the Deutsche Forschungsgemeinschaft through the Excellence 
cluster ``Origins''. R.B. acknowledges support by the Severo Ochoa 2020
research programme of the Instituto de Astrof\'{\i}sica de Canarias.

We acknowledge the use of images from the DSS imaging survey provided by NASA/IPAC,
the PanSTARRS Digital Sky Survey provided by the University of Hawaii and 
the Space Telescope Science Institute and the DESI Legacy Imaging Surveys.
 The DESI Legacy Surveys consist of three individual and complementary projects: the Dark Energy 
Camera Legacy Survey (DECaLS; Proposal ID 2014B-0404; PIs: David Schlegel and Arjun Dey), 
the Beijing-Arizona Sky Survey (BASS; NOAO Prop. ID 2015A-0801; PIs: Zhou Xu and Xiaohui Fan), 
and the Mayall z-band Legacy Survey (MzLS; Prop. ID 2016A-0453; PI: Arjun Dey). 
DECaLS, BASS and MzLS together include data obtained, respectively, at the Blanco 
telescope, Cerro Tololo Inter-American Observatory, NSF’s NOIRLab; the Bok telescope, 
Steward Observatory, University of Arizona; and the Mayall telescope, Kitt Peak National 
Observatory, NOIRLab. The Legacy Surveys project is honored to be permitted to conduct 
astronomical research on Iolkam Du’ag (Kitt Peak), a mountain with particular 
significance to the Tohono O’odham Nation.

\end{acknowledgements}

\appendix
\section{Properties of the observed low redshift clusters}

In this section we describe the properties of the low redshift clusters
($z < 0.25$) which appear in the observation fields of the target clusters.

\begin{table}[h]
\caption{Cluster properties of RXCJ1310.4+2151, RXCJ1310.9+2157 and RXCJ1311.7+2201}
\begin{small}
\begin{center}
\begin{tabular}{lccc}
\hline
{\rm name} & {\rm R..1311+2201} & {\rm RXCJ1414+2711}  & {\rm RXCJ2116-0306} \\
\hline
{\rm RA }       &   13~11~46.4  & 14~14~22.1~ &  21~16~10.7 \\
{\rm DEC }      & +22~01~37   & +27~11~26 & -03~06~17 \\
{\rm spec. z$^{a}$}  &  0.1707 {\rm (8)} & 0.1619  {\rm (11)} & 0.2252 {\rm (3)} \\
{\rm BCG z$^{b}$}    &  0.1715 & 0.1578  & 0.2255        \\         
{\rm X-ray z}        &  $0.18\pm 0.03$ & 0.16 $\pm 0.04$ & 0.23 $\pm 0.03$\\
\hline
 F$_X$       & $5.9 \pm 0.5$ & $3.3 \pm 0.4$  &  $ 2.5 \pm 0.15$ \\
 L$_X$       & $0.43 \pm 0.04$  &  0.22 $\pm 0.025$ & $ 0.35  \pm 0.02$\\
 T$_X$       & $3.25     \pm0.5$  & $2.25 \pm 0.4$ & $ 3.1 \pm 0.4$ \\
{\rm M$_{gas}$}   & 1.6 & 0.70   &   1.6     \\
 r$_c$        & 0.22  & 0.13 & 0.77 \\
 $\beta$       & 0.36 &  0.40 & 0.54  \\ 
\hline
\end{tabular}
\end{center}
\end{small}
{{\bf Notes} The meaning of the parameters is the same as in Table~\ref{tab3}.\\
$^{a)}$ The integers in brackets give the number of galaxies available 
for the cluster redshift determination.\\ 
$^{b)}$ The redshift of the BCG of  RXCJ1311.7+2201 
is from the 2MASS galaxy survey \citep{Bil2014}, 
the redshift of the BCG from SDSS data release 6 (2007) for RXCJ1414.3+2711, 
and the redshifts of the BCG of RXCJ2116.1-0306 from the 2MASS redshift survey
\citep{Bil2014}.
} 
\label{tab10}
\end{table}

\begin{table*}[h]
\caption{Results of different mass estimates for the low redshift galaxy clusters.}
\begin{small}
\begin{center}
\begin{tabular}{lrrr}
\hline
  {\rm mass } & {\rm 1311} & {\rm 1414N} & {\rm 2116N} \\
\hline
$M_{500}(\beta)$   & 0.8$\pm0.1$   & 0.55$\pm0.07$  & 1.24$\pm0.2$\\
$M_{500}(\beta_{fix})$& 2.0$\pm0.3$ & 1.16$\pm0.15$  & 1.7$\pm0.3$ \\
$M_{500}(L_X)$     & 1.8$\pm0.7$   & 1.2$\pm0.5$    & 1.5$\pm 0.6$\\
$M_{500}(T_X)$     &  2.0$\pm0.5$  & 1.15$\pm0.3$   & 1.8 $\pm0.35$ \\
$M_{500}(M_{gas})$ & 1.85$\pm0.3$  & 0.92$\pm0.16$  & 1.8$\pm0.3$ \\
$M_{500}(Y_X)$    & 1.8$\pm0.3$   & 0.91$\pm0.14$   & 1.7$\pm0.22$ \\
$M_{dyn}$         &              &                &         \\
\hline
$M_{500}$        & 1.9$\pm0.3$   & 1.06$\pm0.14$  & 1.7$\pm0.22$ \\
$r_{500}$        & 4.63          & 4.01           & 3.60 \\
\hline
\end{tabular}
\end{center}
\end{small}
{{\bf Notes:} The meaning of the rows are the same as for Table~\ref{tab9}}
\label{tab11}
\end{table*}

\subsection{RXCJ1311.7+2201}

\begin{figure}[h]
   \includegraphics[width=\columnwidth]{RXCJ1311.7+2201_ovl.ps}
\caption{X-ray surface brightness contours overlayed on 
an optical PanSTARRS image for RXCJ1311.7+2201. 
}\label{fig7}
\end{figure}

Its X-ray flux of $F_X \sim 1 \times 10^{-12}$ 
erg s$^{-1}$ cm$^{_2}$ is below the flux-limit
for the {\sf CLASSIX} survey and thus not contained in its catalog. 
But as a cluster well detected in the ROSAT All-Sky Survey, it receives the name, 
RXCJ1311.7+2201. 
This galaxy cluster, which was identified in the {\sf NORAS} survey \citep{Boe2000},
has a redshift of $z = 0.1707$. It was also detected 
in the SDSS \citep{Koe2007,Hao2010,Wen2010}. Table~~\ref{tab10} provides 
information on redshift and X-ray properties of the clusters. An X-ray image
of the cluster is shown in Fig~\ref{fig4} on the left hand side. 
The cluster displays an extension from the compact center towards the
northwest. Fig.~\ref{fig7} presents an overlay of X-ray contours on an 
optical PanSTARRS image. We note a bright lenticular BCG at the peak 
of the X-ray emission.  

In the analysis of the X-ray spectra in different circles and rings at radii,
$r < 0.67$, $< 1.5$, $< 2.5$ and $r = 1 - 2$ arcmin
($< 117, < 262, < 436, 175 - 349$ kpc), we find temperatures of
$T_X = 3.25 \pm 0.6$, $3.3 \pm 0.5$, $3.2 \pm 0.5$ and $2.9 \pm 0.7$ keV, respectively.
A temperature profile with a polytropic index of 1.2 or larger would be
inconsistent with the data. Cluster properties derived from these data are
given in Table~\ref{tab10}. The different mass estimates listed in Table~\ref{tab11}
give consistent results with a final mass of $M_{500} = 1.9 (\pm 0.3) \times
10^{14}$ M$_{\odot}$. The system shows a central 
cooling time of 5.6 Gyr and has thus no cooling core.

\subsection{RXCJ1414.3+2711}

The cluster RXCJ1414.3+2711, which was noted during the {\sf NORAS II}
survey was also found in the Northern Sky Optical Cluster
Survey \citep{Lop2004}. An x-ray image in the 0.5 to 2 keV energy band
of the field of the target cluster RXCJ1414.3+2711 is shown in Fig.~\ref{fig10}.
The cluster has a central component and an extension towards
the south-east. Table~~\ref{tab10} provides information on redshift
and X-ray properties of the clusters.

In the analysis of the X-ray spectra 
we find for $r < 0.5$, $r < 1$, $< 2$ and $r = 1 - 2$ arcmin
($< 84, < 167, < 334, 167 - 334$ kpc), 
temperatures of $T_X = 2.3 \pm 0.6$, $2.4 \pm 0.5$, $2.0 \pm 0.5$ 
and $1.95 \pm 0.7$, respectively. 

\begin{figure}[h]
   \includegraphics[width=\columnwidth]{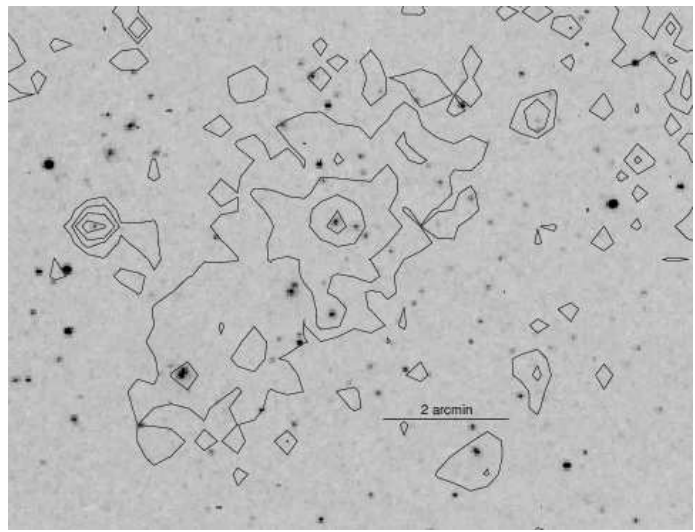}
\caption{X-ray surface brightness contours overlayed on 
an optical DSS image for RXCJ1414.3+2711.
}\label{fig12}
\end{figure}

X-ray contours
superposed on a DSS image for RXCJ1414.3+2711 are shown in
Fig.~\ref{fig12}. This cluster also shows a BCG at the X-ray maximum
and we observe two bright galaxies marking the extension towards
the south-east. The redshifts for the three bright galaxies are from 
north to south: $z = 0.1579$ (SDSS DR 6), $z = 0.1615$ \citep{Bil2014},
and $z = 0.1621$  \citep{Bil2014}.

The mass estimates span a small range of
$M_{500} = 0.91 - 1.2 \times 10^{14}$  M$_{\odot}$, which are all consistent
with the finally adopted value of 
$M_{500} =  1.06 (\pm 0.16)\times 10^{14}$  M$_{\odot}$ (see Table~\ref{tab11}). 
The extension is, however, not included in this mass estimate.

\subsection{RXCJ2116.1-0306}

The foreground cluster RXCJ2116.1-0306, with its center about
three arcmin north of the distant target cluster RXCJ2116.2-0309
is quite extended on the sky and shows some elongation in the north-south
direction. Table~~\ref{tab10} provides information on redshift
and X-ray properties of the clusters. The cluster can be seen in the X-ray 
image given in Fig.~\ref{fig17}.

Fig.~\ref{fig18} shows an image of the X-ray surface brightness 
contours overlayed on an optical i-band image from the
PanSTARRS survey and a colour image from the Legacy Imaging Surveys. 
The cluster RXCJ2116.1-0306 has a BCG at its center.

For RXCJ2116.1-0306 we obtained 
temperatures of $3.15\pm 0.5$, $3.1\pm 0.5$, $2.7\pm 0.6$, and $3.2\pm 0.8$ kev
in regions of $r < 0.5$, $< 1$, $< 1.67$ and $r = 0.5 - 1$ arcmin
($< 109, < 217, < 363, 109 - 217$ kpc), respectively.
The cluster shows a central cooling time of the ICM of about 3 Gyr, which can be 
taken as a sign of a moderate cooling core.

The mass estimates are confined
to a range of $M_{500} = 1.5 - 1.8 \times 10^{14}$  M$_{\odot}$. All estimates are
consistent with the finally adopted mass value of  
$M_{500} =  1.7 (\pm 0.3)\times 10^{14}$  M$_{\odot}$ (see Table~\ref{tab11}).

\section{The variable X-ray source, XMMUJ2116.2-0311}

\begin{figure}[h]
   \includegraphics[width=\columnwidth]{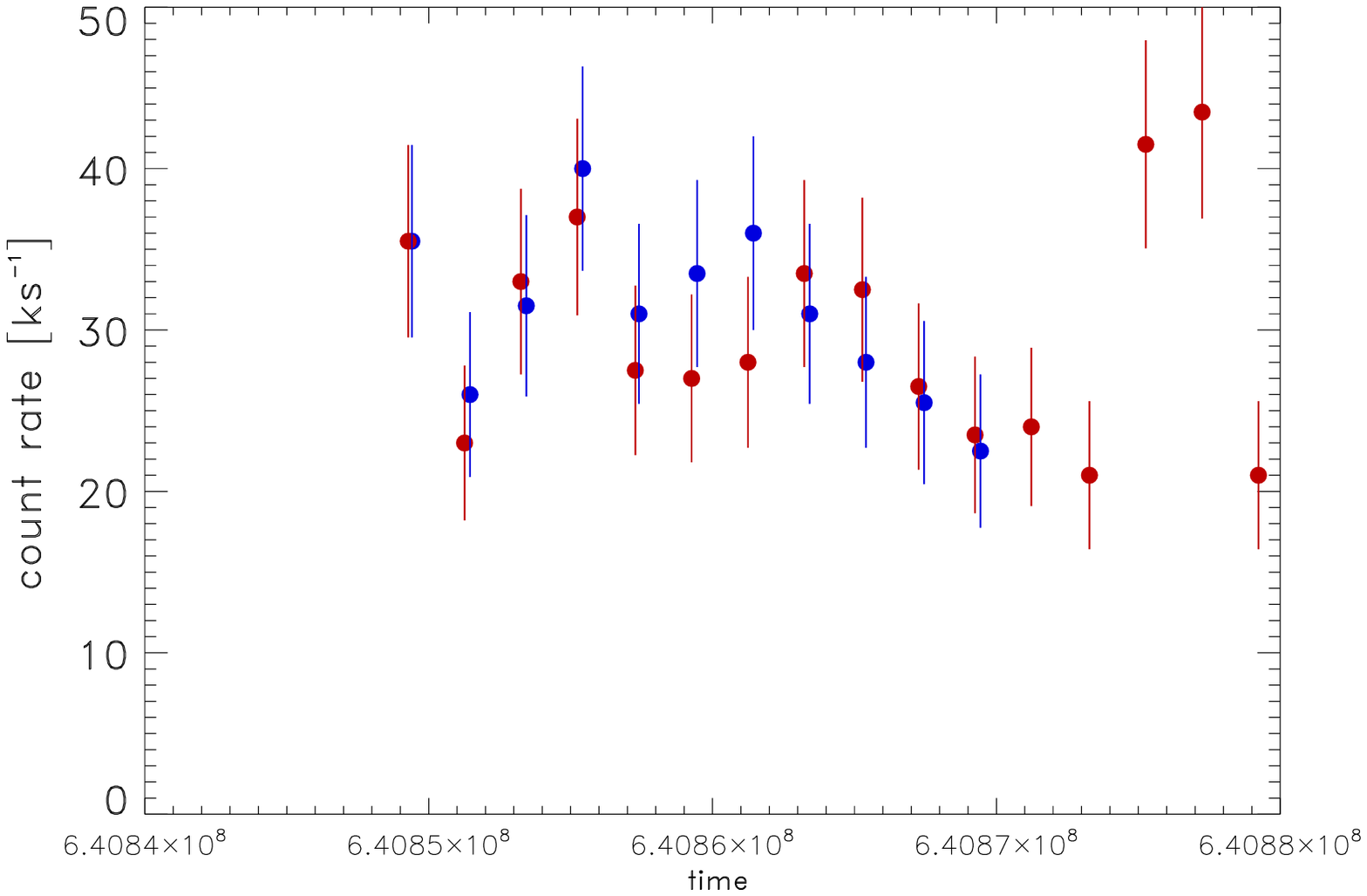}
   \includegraphics[width=\columnwidth]{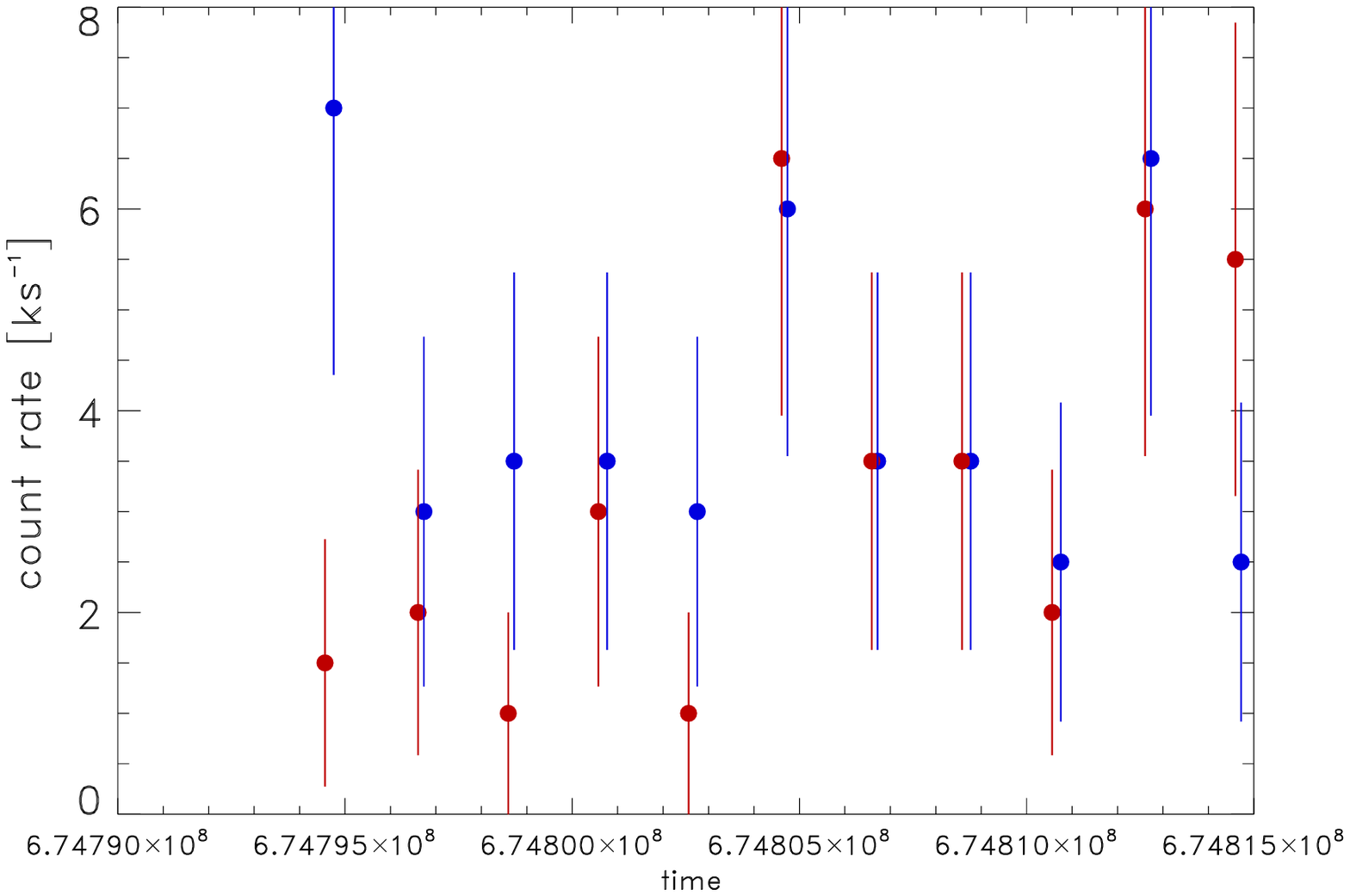}
\caption{Light curves of XMMUJ2116.2-0311 from the two XMM-Newton observations 
in 2018 (top) and 2019 (bottom) recorded by the MOS 1 (blue) and MOS 2 (red)
detectors.
The count rates are for the 0.5 to 2 keV energy 
band. The bin size is 2000 s. The data points for MOS 2
are slightly displaced for better visibilioty.
}\label{figA1}
\end{figure}

The X-ray source XMMUJ2116.2-0311 is covered by the two observations
0803410701 (obs. A) and 0841900601 (Obs. B) 
from 23.4. 2018 and 21.5.2019, respectively.
Fig.~\ref{figA1} shows light curves of the source in the two
epochs of observation in the 0.5 to 5 keV band in bins of 2 ks
as recorded by the MOS 1 and 2 detectors. The PN detector
data cover shorter time intervals.
The extraction region for the count rate was 30 arcsec.
After background subtraction the mean count rate 
corrected for missing flux for a point source
in the extraction aperture in the first
observation is 29.8 ($\pm 1.0$) ks$^{-1}$ and in the second observation
3.6 ($\pm 0.3$) ks$^{-1}$, a change by a factor of 8.3. 
The corresponding 0.5 to 2 keV flux is 
1.86 ($\pm 0.06) \times 10^{-13}$ erg s$^{-1}$ cm$^{-2}$
and 2.1 ($\pm 0.2) \times 10^{-14}$ erg s$^{-1}$ cm$^{-2}$.
The optical image of the counterpart appears as a point-like source
in the PanSTARRS image (Fig.~\ref{fig18}).
The X-ray source is most probably an AGN which changed its accretion
rate or went through a special accretion event in 2019. 
Assuming that the two detections of the source are associated to  
temporary and now fading X-ray emission the source can also be interpreted
as a tidal disruption event of a star accreted by a supermassive 
black hole, e.g. \citet{Kom1999, Sax2021,Saz2021}. The observation
of the sky position in the ROSAT All-Sky Survey 
with an exposure time of 445 s does 
not help to determine if this is a temporary source, since in the 
presence of the extended cluster emission we can only derive an 
upper limit for the flux of about $5 \times 10^{-13}$ erg s$^{-1}$ 
cm$^{-2}$, insufficient to rule out the presence of the source in 
the ROSAT Survey.    


\begin{thebibliography}{}

\bibitem[Appenzeller et al. (1998)]{App1998}
Appenzeller, I., Thiering, I., Zickgraf, F. -J. et al., 1998, ApJS,117,319

\bibitem[Arnaud et al. (2005)]{Arn2005}
Arnaud, M., Pointecouteau, E., Pratt, G.W., 2005, A\&A, 441, 893 

\bibitem[Barrena  et al.(2022)]{Bar2021}
Barrena, R., B\"ohringer, H. \& Chon G., 2022, (A\&A in press), arXiv2205.05597  

\bibitem[Bilicki et al. (2014)]{Bil2014}
Bilicki, M., Jarrett, T.H., Peacock, J.A., et al., 2014, ApLS, 199, 34

\bibitem[B\"ohringer et al. (2000)]{Boe2000}
B\"ohringer, H., Voges, W., Huchra, J.P., et al., 2000, ApJS, 129, 435

\bibitem[B\"ohringer et al. (2004)]{Boe2004}
B\"ohringer, H., Schuecker, P., Guzzo, L., et al., 2004, A\&A, 425, 367

\bibitem[B\"ohringer et al. (2013)]{Boe2013}
B\"ohringer, H., Chon, G., Collins, C.A., et al., 2014, A\&A, 555, A30

\bibitem[B\"ohringer et al. (2014)]{Boe2014}
B\"ohringer, H., Chon, G., Collins, C.A., et al., 2014, A\&A, 570, A31

\bibitem[B\"ohringer et al. (2016)]{Boe2016}
B\"ohringer, H., Chon, G., Kronberg, P.P., 2016, A\&A, 596, 22 

\bibitem[B\"ohringer et al. (2017)]{Boe2017}
B\"ohringer, H., Chon, G., Retzlaff, J., et al., 2017, AJ, 153, 220 

\bibitem[Borgani et al. (2004)]{Bor2004}
Borgani, S., Murante, G., Springel, V., et al. 2004, MNRAS, 348, 1078

\bibitem[Bouwens et al. (2014)]{Bou2014}
Bouwens, R.J., Bradley, L., Zitrin, A. et al., 2014, ApJ, 795, 126

\bibitem[Cavaliere \& Fusco-Femiano (1976)]{Cav1976}
Cavaliere, A. \& Fusco-Femiano, R., 1976, A\&A, 49, 137 

\bibitem[Chon \& B\"ohringer (2017)]{Cho2017}
Chon, G., \& B\"ohringer, H., 2017, A\&A, 606, L4

\bibitem[Croston et al. (2008)]{Cro2008}
Croston, J.H., Pratt, G.W., Böhringer, H., et al., 2008, A\&A, 48, 431 

\bibitem[De Grandi \& Moldendi (2004)]{DeG2004}
De Grandi, S. \& Moldendi, S., 2004, arXiv:astro-ph/0407392 

\bibitem[Dickey (1990)]{}
Dickey, J.M. \& Lockman, F.J., 1990, ARA\&A, 28, 215

\bibitem[Ellis (2014)]{Eli2014}
Ellis, R.S., 2014, arXiv1411.3330

\bibitem[Feretti et al. (2012)]{Fer2012}
Feretti, L., Giovannini, G., Govoni, F., et al., 2012, A\&ARv, 20, 54 

\bibitem[Fischer et al. (1998)]{Fis1998}
Fischer, J.-U., Hasinger, G., Schwope, A. D., et al., 1998, AN, 319, 347 

\bibitem[Fujimoto et al. (2016)]{Fuj2016}
Fujimoto, S., Ouchi, M., Ono, Y., et al., 2016, ApJS, 222, 1

\bibitem[Gal et al. (2003)]{Gal2003}
Gal, R.R., de Carvalho, R.R., Lopes, P.A.A., 2003, AJ, 125, 2064

\bibitem[Guzzo (2009)]{Guz2009}
Guzzo, L., Schuecker, P., B\"ohringer, H., et al., 2009, A\&A, 499, 357

\bibitem[Hao et al. (2010)]{Hao2010}
Hao, J., McKey, T.A., Koester, B.P., et al., 2010, ApJS, 191, 54

\bibitem[Koester et al. (2007)]{Koe2007}
Koester, B.P., McKey, T.A., Annis, J., et al., 2007, ApJ, 660, 239 

\bibitem[Komossa \& Greiner (1999)]{Kom1999}
Komossa, S. \& Greiner, J., 1999, A\&A, 349, L45

\bibitem[Kravtsov et al. (2006)]{Kra2006}
Kravtsov, A.V., Vikhlinin, A., Nagai, D., 2006, ApJ, 650, 128 

\bibitem[Lopes et al. (2004)]{Lop2004}
Lopes, P.A.A., de Carvalho, R.R., Gal, R.R., et al., 2004, AJ, 128, 1017

\bibitem[Mantz et al. (2014)]{Man2014}
Mantz, A.B., Allen, S.W., Morris, R.G., 2014, MNRAS, 440, 2077 

\bibitem[Mazzotta et al. (2004)]{Maz2004}
Mazzotta, P., Rasia, E., Moscardini, L., et al., 2004, MNRAS, 354, 10

\bibitem[McLeod et al.(2016)]{McL2016}
McLeod, D.J.,  McLure, R.J., Dunlop, J.S., 2016, MNRAS, 459, 3812

\bibitem[Nagai (2006)]{Nag2006}
Nagai, D., 2006, ApJ, 650, 538

\bibitem[Nagai et al. (2007)]{Nag2007}
Nagai, D., Kravtsov, A., Vikhlinin, A., 2007, ApJ, 668, 1 

\bibitem[PLANCK Collaboration (2016)]{Pla2016}
PLANCK Collaboration 2016, A\&A,594, A27

\bibitem[Pratt et al. (2007)]{Pra2007}
Pratt, G.W., Böhringer, H., Croston, J.H., et al. 2007, A\&A, 461, 71 

\bibitem[Pratt et al. (2009)]{Pra2009}
Pratt, G.W., Croston, J.H., Arnaud, M., et al., 2009, A\&A, 498, 361

\bibitem[Rasia et al. (2012)]{Ras2012}

Rasia, E., Meneghetti, M., Martino, R., et al., 2012, NJph, 14, 4269 

\bibitem[Rozo et al. (2015)]{Roz2015}
Rozo, E., Rykoff, E.S., Becker, M., et al., 2015, MNRAS, 453, 38 

\bibitem[Sarazin (2002)]{Sar2002}
Sarazin, C.L., 2002, in {\it Merging Processes in Galaxy Clusters},
L. Ferretti, I.M. Gioia, G. Giovannini (eds.) Kluver Acad. Publ.,
p. 1

\bibitem[Saro et al. (2013)]{Sar2013}
Saro, A., Mohr, J.J., Bazin, G., et al., 2013, 772, 47

\bibitem[Saxton et al. (2021)]{Sax2021}
Saxton, R., Komossa, S., Auchetti, K., et al., 2021, Space Sci Rev, 217, 18

\bibitem[Sazonov et al. (2021)]{Saz2021}
Sazonov, S., Gilfanov, M., Medvedev, P., et al., 2021, MNRAS, 508. 3820

\bibitem[Sunyaev \& Zeldovich (1972)]{Sun1972}
Sunyaev, R.A. \& Zeldovich, Ya.B., 1972, CoASP, 4, 173S  

\bibitem[Tr\"umper (1993)]{}
Tr\"umper, J., 1993, Science, 260, 1769

\bibitem[Vikhlinin et al. (2006)]{Vik2006}
Vikhlinin, A.; Kravtsov, A.; Forman, W., et al., 2006, ApJ, 640, 691

\bibitem[Vikhlinin et al. (2009)]{Vik2009}
Vikhlinin, A., Kravtsov, A.V., Burenin, R.A. et al., 2009, ApJ, 692,1060 

\bibitem[Voges (1999)]{}
Voges, W., Aschenbach, B., Boller, T., et al. 1999,  A\&A, 349, 389

\bibitem[Wen et al. (2009)]{Wen2009}
Wen, Z.L., Han J.L.,  \& Liu, F.S., 2009, ApJS, 183, 197; 2010, ApJS, 187, 272

\bibitem[Wen et al. (2010)]{Wen2010}
Wen, Z.L., Han, J.L., Liu, F.S., 2010, ApJS, 187, 272

\bibitem[Wen et al. (2012)]{Wen2012}
Wen, Z.L., Han, J.L., Liu, F.S., 2012, ApJS, 199, 34

\bibitem[Wen \& Han (2015)]{Wen2015}
Wen, Z.L. \& Han, J.L., 2015, ApJ, 807, 178

\bibitem[Zhang et al. (2008)]{Zha2008}
Zhang, Y.-Y., Finoguenov, A., Böhringer, H., et al. 2008, A\&A, 482, 451

\end{thebibliography}
\end{document}